\documentclass[review]{elsarticle}
\usepackage[commentmarkup=todo,authormarkup=none]{changes}

\biboptions{sort&compress,numbers}
\usepackage[margin=1in]{geometry}
\usepackage{float}
\usepackage[colorlinks=false]{hyperref}\hypersetup{pdfborder={0 0 0}}
\usepackage{miller}
\usepackage{bm,amsmath,amssymb,amsthm}
\usepackage{textcomp}
\usepackage{gensymb}
\usepackage{enumitem}
\usepackage{mathalfa}
\usepackage{cancel}
\usepackage{subfigure}\usepackage{graphicx}
\usepackage[margin=1in]{geometry}

\definecolor{C0}{HTML}{1F77B4}
\definecolor{C1}{HTML}{FF7F0E}
\definechangesauthor[color=C0]{R1} 
\definechangesauthor[color=C1]{R2} 

\DeclareUnicodeCharacter{03B3}{$\gamma$}
\journal{Acta Materialia}

\begin{document}

\begin{frontmatter}
\title{On the Variability of Grain Boundary Mobility in the Isoconfigurational Ensemble}
\author[cmu]{Anqi Qiu}
\ead{anqiq@andrew.cmu.edu}
\cortext[cor1]{Corresponding author}
\author[gmu]{Ian Chesser}
\ead{ichesser@gmu.edu}
\author[cmu]{Elizabeth Holm
\corref{cor1}}\ead{eaholm@andrew.cmu.edu}

\address[cmu]{Department of Materials Science and Engineering, Carnegie Mellon University, Pittsburgh, PA 15213}
\address[gmu]{Department of Physics \& Astronomy, George Mason University, Fairfax, VA 22030}

\begin{abstract}

Recent grain growth experiments have revealed that the same type of grain boundary can have very different mobilities depending on its local microstructure. In this work, we use molecular dynamics simulations to quantify uncertainty in the reduced mobility of curved grain boundaries for different types of boundary conditions and over a range of initial velocity seeds. We consider 
cylindrical island grains in Ni with a [001] tilt axis as a model system.  Unexpectedly, we find large variation in the reduced mobility of curved grain boundaries depending on both the imposed constraints and the initial velocity distribution. We perform a dynamic propensity analysis inspired from studies of glass forming liquids to analyze sources of variation in reduced mobility. Our work highlights the significant impact of initial velocity distributions on grain boundary motion which has not been analyzed in prior work.

\end{abstract}

\end{frontmatter}



\section{Introduction}

Grain boundaries are interfaces that occur at the intersections of grains of the same phase with different orientations \cite{sutton_interfaces_2006}. Due to having more disordered local structures than perfect crystals, they possess much lower energy barriers to atomic rearrangement. Therefore, grain boundaries tend to migrate to reduce system energy, leading to microstructural evolution.  

There has been increasing awareness that grain boundary motion is not deterministic, but is inherently statistical \cite{race_quantifying_2015}, and that grain boundary mobility may not be an intrinsic property independent of driving force as was widely believed \cite{chen_grain_2020}. Previous works have shown that grain boundary structure is an ensemble of multiple metastable states that form a wide energy band, not a single configuration with definite energy \cite{han_grain-boundary_2016}. However, the effects of different initial velocity distributions on grain boundary migration, an additional source of possible variability, have not been explored in detail. 

The isoconfigurational ensemble is the combined set of different motion trajectories of the same initial structure with different initial momenta sampled from the Maxwell-Boltzmann distribution at the desired temperature \cite{widmer-cooper_how_2004,widmer-cooper_study_2007}. It has been widely applied in the studies of atomic motions in amorphous solids and supercooled liquids, but its application in crystalline solids has never been a topic of study. In liquids and amorphous solids, atomic motions are not well constrained by neighboring atoms, so relevant molecular dynamics (MD) studies take into account the effects of initial momenta distributions, by performing multiple simulation runs with different initial velocities sampled from the same Maxwell-Boltzmann velocity distribution for the same initial configuration. By averaging over the isoconfigurational ensemble to remove the noises imposed by the initial velocities, the collective dynamics of atoms and their relationships to the initial structure can be more clearly shown. In crystalline solids, atomic motions are more constrained by the crystalline lattices and considered to be predictable, and very few, most often only a single simulation is performed for a given configuration. Grain boundaries have been shown to exhibit the dynamics of glass-forming liquids \cite{zhang_grain_2009}, with structures similar to amorphous solids that are less constrained by neighboring atoms than in crystalline lattices. Therefore, research on grain boundary motion may benefit from the adoption of the isoconfigurational ensemble method.

We implement the well-known shrinking cylindrical grain model \cite{zhang_determination_2006,upmanyu_simultaneous_2006,trautt_grain_2012,molodov_grain_2015,french_molecular_2022} in the isoconfigurational ensemble, using MD simulations. The shrinking cylindrical grain method has been widely applied in the study of grain boundary motion, but none of the previous works have taken into account the effects of initial velocity distributions on grain boundary motion. The cylindrical grain boundary samples a full spectrum $(0-360\degree)$ of different boundary inclinations and being curved, is geometrically close to realistic grain boundaries. Real materials are usually composed of networks of interconnecting and interacting curved grain boundaries, making it difficult to examine the motions of individual grain boundaries. The isolated cylindrical grain boundary is not affected by the motions of other boundaries and when heated, will shrink spontaneously under curvature driving force, needing no external driving force. Therefore, it is ideal for investigating grain boundary motion. 

This paper is structured as follows. In Section \ref{sec:methods}, we elucidate the methodology used in this study, including simulation setup, grain segmentation, grain boundary mobility calculation, and grain rotation measurement. In Section \ref{sec:results}, we highlight the significant impact of initial velocity distributions on grain boundary motion and associated grain rotation under various boundary conditions. In Section \ref{sec:analysis}, we present possible explanations to identify the source of the variability that we observed. All the main conclusions are summarized in Section \ref{sec:conclusions}.

\section{Methods}
\label{sec:methods}

\subsection{Simulation setup}
All simulations are performed using the Large-scale Atomic/Molecular Massively Parallel Simulator (LAMMPS) \cite{plimpton_fast_1995} in the isothermal-isobaric (NPT) ensemble. Ni is a typical Face Centered Cubic (FCC) metallic material for which grain boundary properties have been widely studied, notably in the Olmsted survey in which the properties of 388 boundaries are investigated \cite{olmsted_survey_2009,olmsted_survey_2009-1}. The initial bicrystal structures are constructed by creating a simulation box the size of 60${a_0}$×60${a_0}$×5${a_0}$ (211.2Å×211.2Å×17.6Å) in terms of the Ni lattice parameter ${a_0}$ = 3.52Å, and rotating a cylindrical inner grain with the radius of 15${a_0}$ (53.8Å) located in the center by a defined misorientation angle $\theta$ around the [0 0 1] tilt axis. After the inner grain is rotated, pairs of atoms whose separation distances are within 1.6Å of each other are searched for, and one of them is deleted. The cutoff distance of 1.6Å is used because it results in the lowest grain boundary energy after structural optimization. The cylindrical grain boundary structure is of ${D_{4h}}$ symmetry, so only initial misorientations of 0-45$\degree$ are investigated.

The structures are first optimized by molecular statics using the Polak-Ribiere version of the conjugate descent algorithm \cite{hestenes_methods_1952} at 0 K. After structural optimization, random initial velocities sampled from the Maxwell-Boltzmann distribution are assigned to all mobile atoms, according to the desired temperature. The different initial velocity distributions are indicated by initial velocity seeds in LAMMPS. In order to reduce the internal stress created by the loss of grain boundary free volume during grain shrinkage, the isothermal-isobaric, or NPT ensemble is used, with zero pressure applied in all three Cartesian directions. The system temperature is set to a defined temperature between 700 K and 1400 K, and the timestep is set to 0.003 ps. Periodic boundary conditions are specified in all three Cartesian directions. The Foiles-Hoyt embedded atom method (EAM) interatomic potential for Ni, which has been shown to accurately predict 
stacking fault energy and elastic moduli in real materials \cite{foiles_computation_2006}, is used to represent the interactions between atoms. The same potential was used in various studies, notably the Olmsted survey in which the properties of 388 Ni grain boundaries are explored \cite{olmsted_survey_2009,olmsted_survey_2009-1}. The melting point associated with the potential is 1565 K.
\begin{figure}[h]
    \centering\leavevmode
    \includegraphics[width=1\textwidth]{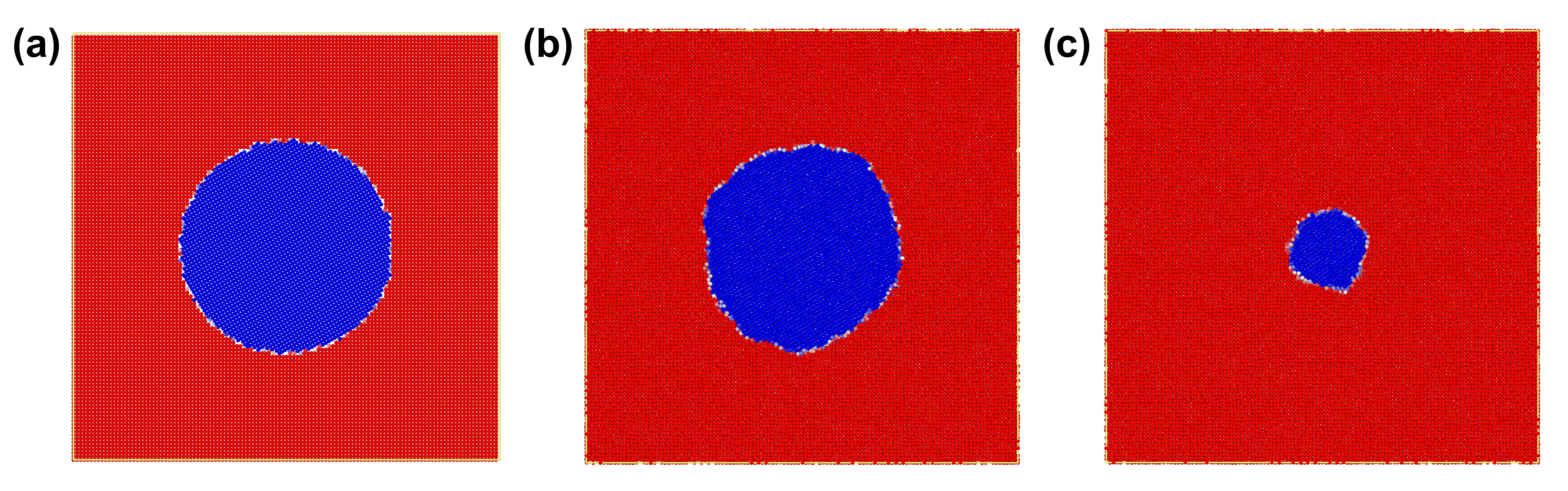}
    \caption{Freely shrinking cylindrical grain with initial misorientation of 30$\degree$  at simulations times of (a) 0 ps, (b) 30 ps, (c) 180 ps}
    \label{fig:S1}
\end{figure}

When the system is heated to a high temperature ($>$ 700 K), the embedded cylindrical grain will shrink spontaneously under the curvature driving force and disappear within 1 ns. Figure \ref{fig:S1} shows snapshots of the simulation box during the shrinking process. The visualizations are performed using the Open Visualization Tool (OVITO) \cite{stukowski_visualization_2009}. The blue atoms represent atoms in the cylindrical grain, red atoms represent atoms in the matrix grain, and atoms at the interface of the two grains are identified as grain boundary atoms. The atoms are colored according to an order parameter, as defined by Schratt and Mohles \cite{schratt_efficient_2020}. During the shrinking process, the inner grain is consistently cylindrical until the grain becomes so small that it disappears, though its shape is constantly changing. 

\subsection{Grain segmentation}
Before any analysis can be performed, it is necessary to identify the atoms in each grain and grain boundaries atoms. In this work, we use an orientation-dependent order parameter, as defined by Schratt and Mohles \cite{schratt_efficient_2020} for grain segmentation. A symmetric orientational order parameter for atom $i$ is defined as:
\begin{equation}\label{1}
\chi_{i} = \frac{1}{N}\sum_{k=1}^{3}[|\psi_k^I(\mathbf{r_i})|^2-|\psi_k^{II}(\mathbf{r_i})|^2]
\end{equation}
where N indicates the normalization factor at $T = 0 K$, the complex functions $\psi_k^I(\mathbf{r_i})$ and $\psi_k^{II}(\mathbf{r_i})$
are used to indicate how closely an atom $i$ matches the perfect orientations of crystals I and II, as described in terms of the three reciprocal lattice vectors $\mathbf{r_1}, \mathbf{r_2}, \mathbf{r_3}$. The order parameter $\chi_i$ is normalized to the range of [-1,+1] at $T = 0 K$, and can exceed the range at higher temperatures, so a cutoff parameter $\eta$ is used. The new order parameter $X$ is defined as:
\begin{equation}\label{2}
X_{i} = \begin{cases}
+1, & \chi_{i} \geq \eta\\
\sin{\frac{\pi \chi_{i}}{2 \eta}}, & -\eta < \chi_{i} < \eta\\ -1, & \chi_{i} \leq -\eta
\end{cases} 
\end{equation}

In our simulations, a cutoff value of $\eta = 0.25$ is used. The matrix grains are assigned $X_{i} = 1$, the cylindrical grain atoms are assigned $X_{i} = -1$, and atoms with $-1 < X_{i} < 1$ are identified as grain boundary atoms. The segmentation results match very well with our observations.

\subsection{Reduced mobility calculation}

 Grain boundary velocity v is assumed to be proportional to the driving pressure P:
\begin{equation}\label{3}
v = MP
\end{equation}

This relationship holds for cases in which the driving pressure is sufficiently small: $P << k_B T/\ohm$ \cite{sutton_interfaces_2006,levy_structure_1969,burke_recrystallization_1952}, where $k_B T$ is the product of the Boltzmann constant and temperature, and $\ohm$ is the atomic volume. The coefficient of proportionality, $M$, is defined as the mobility. In curved boundaries, the driving pressure, also known as the curvature driving force or capillarity driving force, can be defined as the product of the grain boundary stiffness $\Gamma$ and the curvature $\kappa$:
\begin{equation}\label{4}
P = \Gamma \kappa = \frac{\Gamma}{r}
\end{equation}
where the curvature $\kappa$ is the reciprocal of the radius $r$. The grain boundary stiffness $\Gamma$ at a certain point is the combination of the grain boundary free energy per unit area $\gamma$ at that point and its second derivative with respect to the boundary plane inclination, ignoring the higher order terms:
\begin{equation}\label{5}
\Gamma = \gamma + \gamma^{''}
\end{equation}
Curved boundaries tend to migrate toward the centers of curvature. The velocity of curved boundaries driven only by the curvature driving force can be represented as:
\begin{equation}\label{6}
v = M\Gamma\kappa
\end{equation}
In curved grain boundaries, during migration, faceting and grain rotation often occur, making it difficult to determine the grain boundary stiffness. Only the product of the grain boundary mobility $M$ and stiffness $\Gamma$, known as the reduced mobility $M^{*}$, can be directly extracted from grain boundary migration \cite{herring_surface_1999}:

\begin{equation}\label{7}
M^{*} = M\Gamma
\end{equation}
Grain boundary mobility is a key parameter governing grain boundary migration \cite{humphreys_chapter_2004}, and is important to understanding the microstructure and properties of materials. In curved grain boundaries, since it is not straightforward to determine mobility, reduced mobility becomes the key parameter for quantifying grain boundary migration rate. 

Reduced mobility measures the perpendicular motion towards the center of curvature of curved grain boundaries and in our case, can be calculated by tracking the evolution of the number of atoms in the cylindrical grain with time. The change in the number of atoms with time can be represented as:
\begin{equation}\label{8}
N = N_0 - 
\frac{2 \pi h M^{*} t}{\ohm} 
\end{equation}
where $N_0$ is the initial number of atoms in the cylindrical grain, $N$ is the current number atoms in the cylindrical grain, $h$ is the height of the simulation box, and $\ohm$ is the volume of the unit cell. This indicates that the number of atoms in the cylindrical grain, which also corresponds to the volume of the grain, decreases linearly with time, and the reduced mobility is constant. The change in the volume of the cylindrical grain with time can be written as:
\begin{equation}\label{9}
V = V_0 - 
2 \pi h M^{*} t
\end{equation}
In the cases where the volume vs. time plot can be well fitted by linear functions, the reduced mobility can be easily extracted from the slope of the linear plot.

\begin{figure}[h]
    \centering\leavevmode
    \includegraphics[width=1\textwidth]{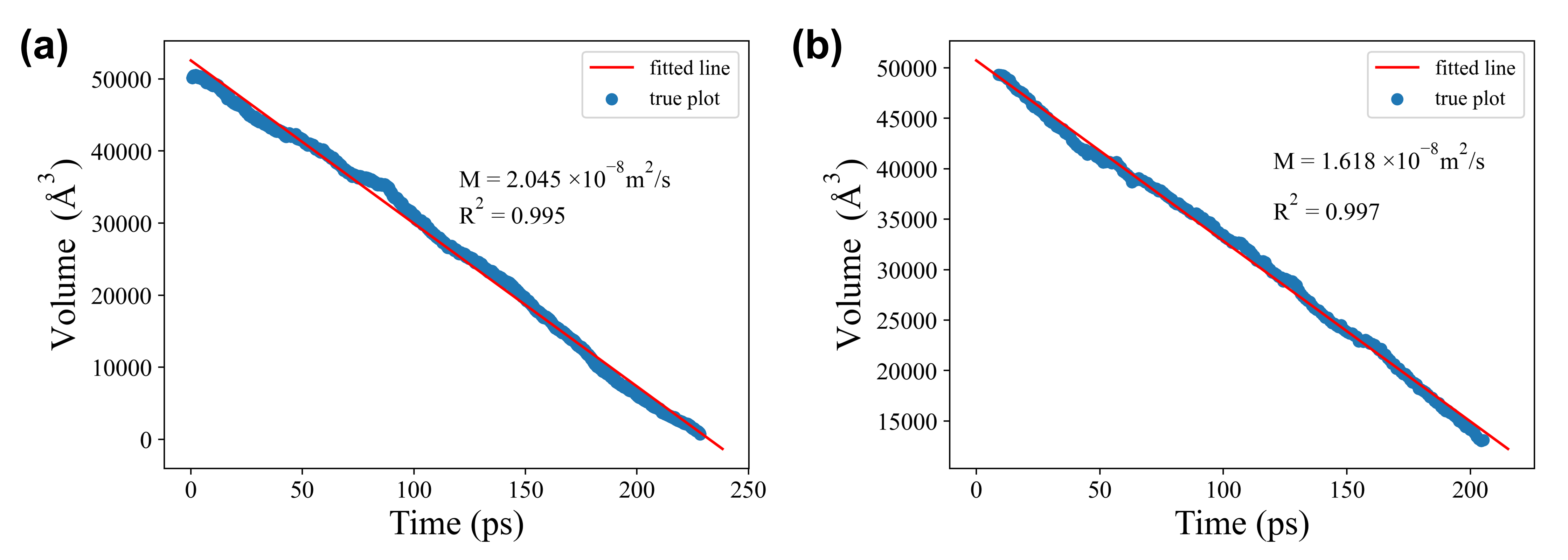}
    \caption{Volume vs. time plot for $\theta = 30\degree$ (a) free (b) fixed boundary at 1000 K}
    \label{fig:S2}
\end{figure}

For each initial misorientation, two different types boundary conditions are applied. The free, or unfixed boundary condition allows the all the atoms in the simulation box to move freely under the capillary driving force. The fixed boundary condition allows all but a small group of atoms in the center of the cylindrical grain to move freely, effectively fixing the core of the cylindrical grain to eliminate grain rotation. For the $\theta = 30\degree$ boundary, for example, under both the free and fixed conditions, the volume vs. time plots are well-fitted to linear functions, as is shown in Figure \ref{fig:S2}. Although there are slight changes in the slopes of the volume vs. time plots throughout the grain boundary migration process, due to the changes in grain boundary structure, we assume the overall reduced mobility to be constant. 

We mainly investigate high angle grain boundaries with initial misorientations of 20$\degree$, 25$\degree$, 30$\degree$, 35$\degree$, 40$\degree$, and 45$\degree$ under free and fixed boundary conditions. Low angle grain boundaries are not investigated in detail because of the frequent occurrence of faceting during migration, causing the motions to be more erratic. Thus reduced mobility is not constant and measurable in low angle grain boundaries. For more information regarding low angle grain boundaries, refer to Supplementary Information.

\subsection{Grain rotation measurement}
The freely moving cylindrical inner grain undergoes rotation during spontaneous shrinkage. Grain rotation can be visualized and measured using fiducial markers, but the method is inaccurate and inefficient, and will not be used extensively for rotation measurement. For more details regarding fiducial markers, see Supplementary Information.

Grain rotation can be reliably measured using quaternions. A quaternion can be represented by four elements:

\begin{equation}\label{10}
\mathbf{q} = q_0 + q_1 \mathbf{i} + q_2 \mathbf{j} + q_3 \mathbf{k}
\end{equation}

, where $q_0$, $q_1$, $q_2$, $q_3$ are real numbers known as the quaternion's w, x, y, and z components, and $\mathbf{i}$, $\mathbf{j}$, and $\mathbf{k}$ are imaginary unit vectors in the Cartesian x, y, and z directions.

In our simulations, the axis of rotation is $\mathbf{w} = [0 0 1]$, and the rotation quaternion for an arbitrary rotation angle of $\alpha$ around $\mathbf{w}$ can be represented as:
\begin{equation}\label{11}
\mathbf{q} = cos(\frac{\alpha}{2}) + sin(\frac{\alpha}{2})[w_x \mathbf{i} + w_y \mathbf{j} + w_z \mathbf{k}] = cos(\frac{\alpha}{2}) + sin(\frac{\alpha}{2}) \mathbf{k}  
\end{equation}

\begin{figure}[h]
    \centering\leavevmode
    \includegraphics[width=0.6\textwidth]{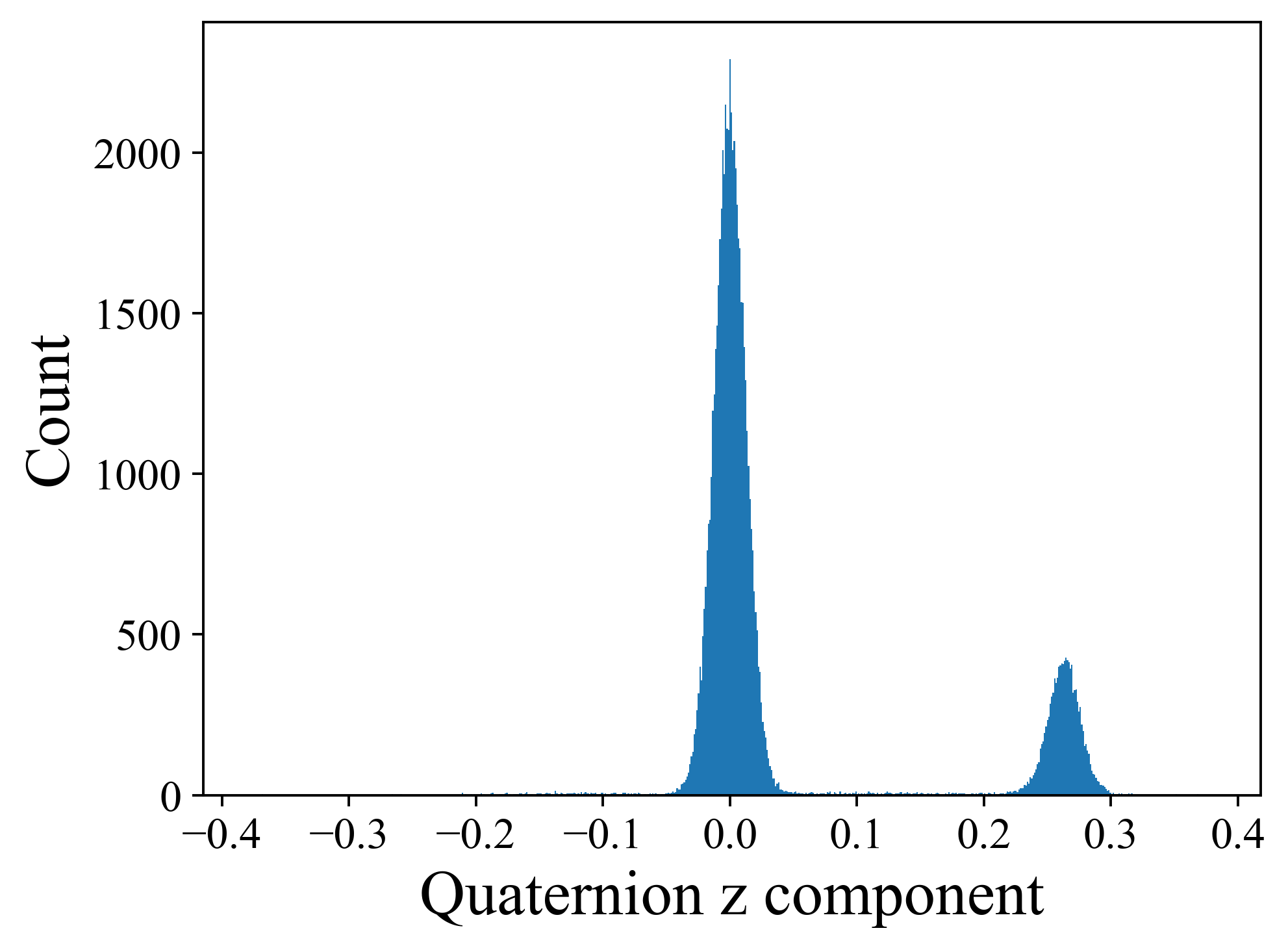}
    \caption{The bimodal distribution of the quaternion z-component for the $\theta = 30\degree$ free boundary at $t=0$}
    \label{fig:bimodal}
\end{figure}

The quaternion z-components of all the atoms in the simulation box form a bimodal distribution, as is shown in Figure \ref{fig:bimodal}. The z-component values of the larger peak are centered around 0, and correspond to atoms in the outer matrix grain. The smaller peak corresponds to atoms in the inner cylindrical grain. By averaging over the z-components of all the atoms in the inner grain, the misorientation angle $\theta$ can be obtained. As the simulation proceeds, the number of atoms in the two peaks changes, but the bimodal distribution maintains, until the inner grain becomes so small that it is no longer distinguishable.

\section{Results and discussions}
\label{sec:results}
\subsection{Reduced mobility}
\begin{figure}[h]
    \centering\leavevmode
    \includegraphics[width=\textwidth]{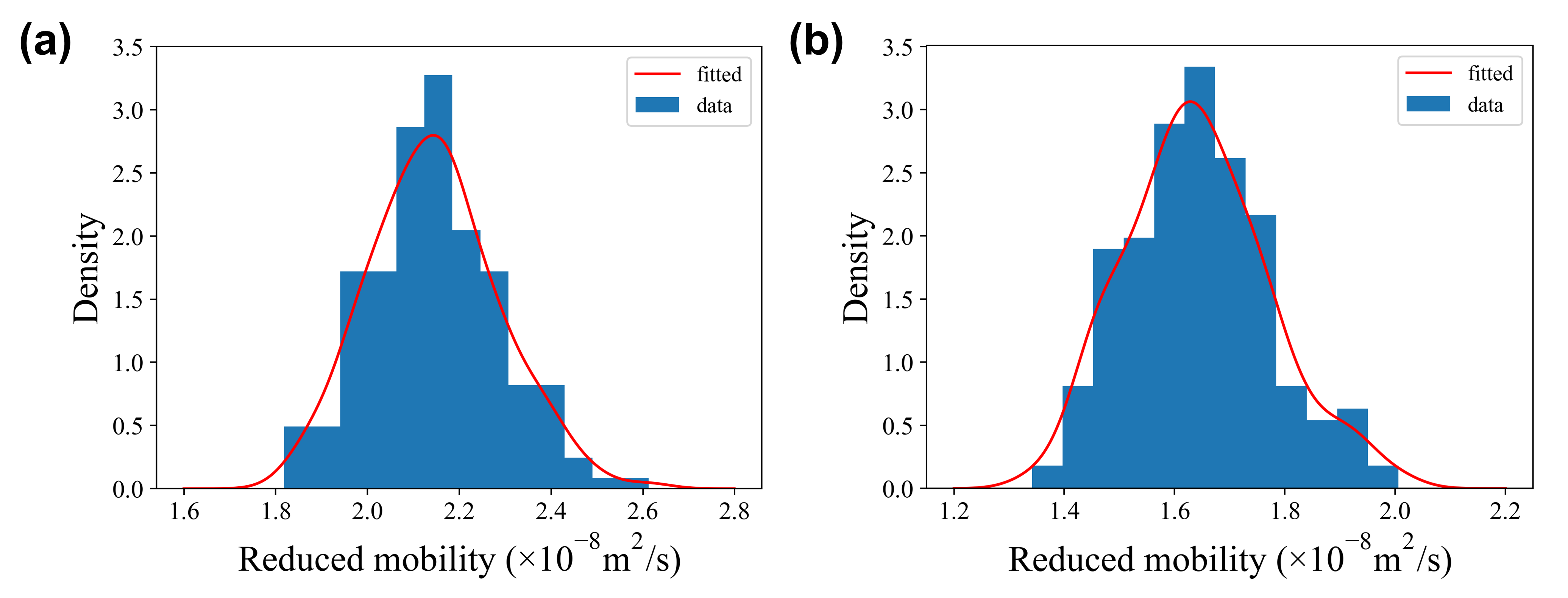}
    \caption{Reduced mobility distribution and corresponding kernel density estimation for $\theta = 30\degree$(a) free (b) fixed boundary at 1000 K}
    \label{fig:gaussian}
\end{figure}

In multiple replicas of the same initial system, the atomic velocities are initialized according to the desired system temperature with random velocities sampled from the Maxwell-Boltzmann distribution, specified by the different initial random velocity seeds in LAMMPS. Previously, the initial velocity distribution has not been considered an important contributing factor in the simulations of grain boundary motion, and in this work, we investigate its effects on grain boundary motion in detail. In our simulations, the same initial structure is shown to have a large variance in reduced mobility. For the $\theta = 30\degree$ free and fixed boundaries at 1000 K, reduced mobilities from 200 different initial velocity seeds are obtained for each boundary. The temperature of 1000 K is selected because it is a temperature at which most boundaries can undergo steady state motion while retaining a large degree of structural order without pre-melting\cite{zhang_grain_2009}. The reduced mobility distributions and their corresponding kernel density estimation (KDE) curves are plotted in Figure \ref{fig:gaussian}. Various normality tests confirm that the two distributions are Gaussian, including the Shapiro–Wilk test, the Kolmogorov-Smirnov test, and Q-Q plot. It is notable that the observed mobilities vary significantly between replicas, with a range of almost $\pm20\%$ of the mean value. This implies that individual measurements of grain boundary mobility may have significant uncertainty.

\begin{figure}[h]
    \centering\leavevmode
    \includegraphics[width=0.6\textwidth]{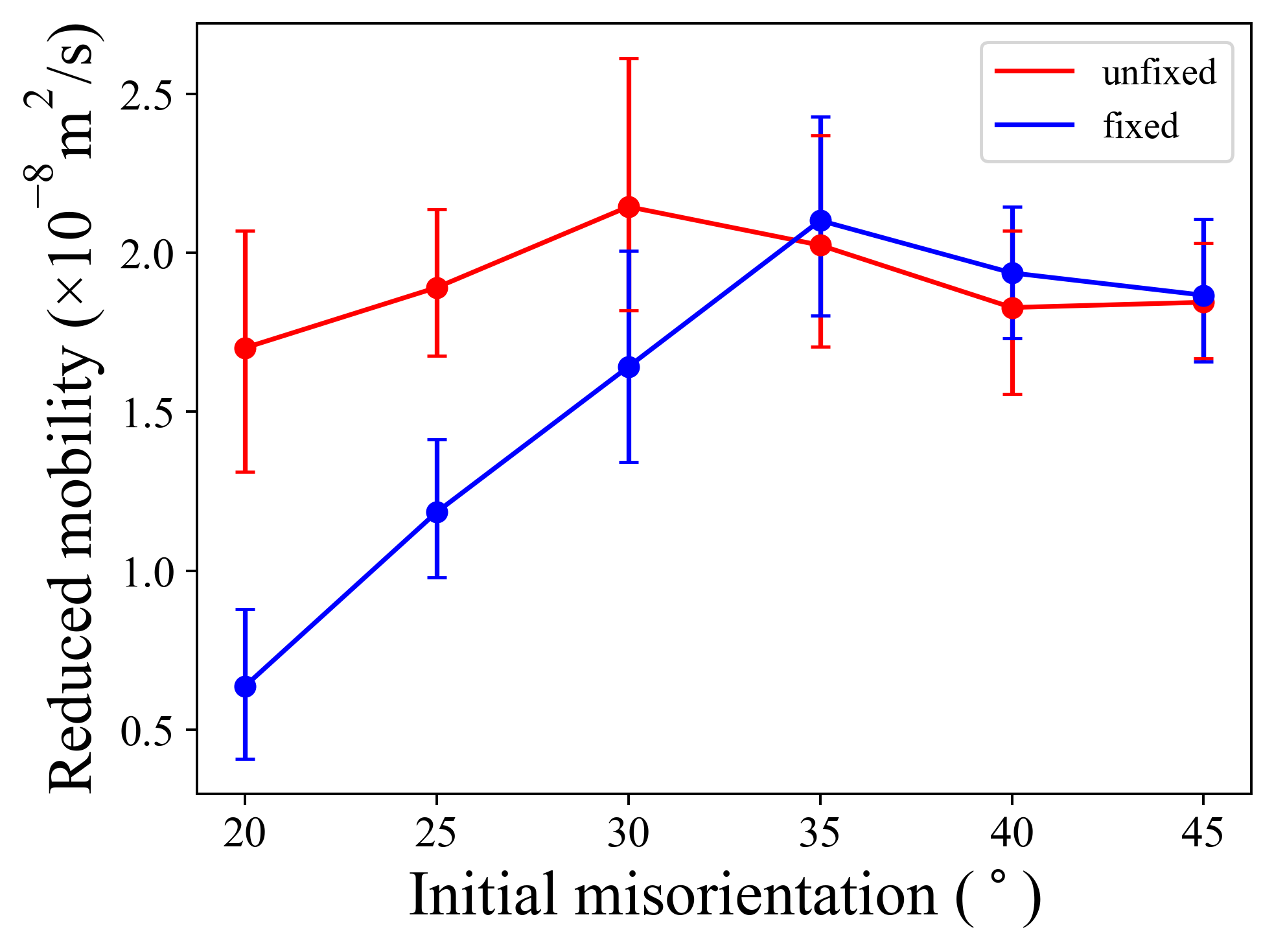}
    \caption{Reduced mobilities of high angle fixed and free boundaries at 1000 K.}
    \label{fig:mobilities-1000K}
\end{figure}

For all the other investigated boundaries, 30 different simulation runs are performed for each boundary, as summarized in Figure \ref{fig:mobilities-1000K}. The reduced mobility datasets for all the investigated grain boundaries passed the same normality tests as the $\theta = 30\degree$ boundaries, indicating that the reduced mobility distributions for the other boundaries are also Gaussian. The 30 simulations are able to give a good estimate of the ensemble average of reduced mobility, as well as the range. In all the investigated cases, the spread in observed mobility values is large and comparable in magnitude. The somewhat wider range of values for the $\theta = 30\degree$ boundaries is attributable to the larger number of observations. This is surprising, considering the only difference between initial states is the assignment of atomic velocity vectors from the same probability distribution.

For cylindrical grains with initial misorientations $< 35\degree$, the average reduced mobilities of the fixed boundaries are significantly lower than their unfixed counterparts. For initial misorientations $\geq 35\degree$, the gaps in reduced mobility between fixed and unfixed boundaries are less significant. The difference in reduced mobility of the fixed and free boundaries is related to grain rotation, as discussed below.

\subsection{Grain rotation}
Grain rotation has been widely observed during grain growth and plastic deformation, and like grain boundary migration, is also a crucial part of microstructural evolution. The rotation of the embedded cylindrical grain has been observed in the shrinking cylinder systems\cite{upmanyu_simultaneous_2006,trautt_grain_2012,french_molecular_2022}, in agreement with Cahn and Taylor's model for shear-coupling induced grain rotation\cite{cahn_unified_2004}. In our simulations, the differences in reduced mobility for the fixed and freely shrinking boundaries can be attributed to grain rotation. 

\begin{figure}[h]
    \centering\leavevmode
    \includegraphics[width=\textwidth]{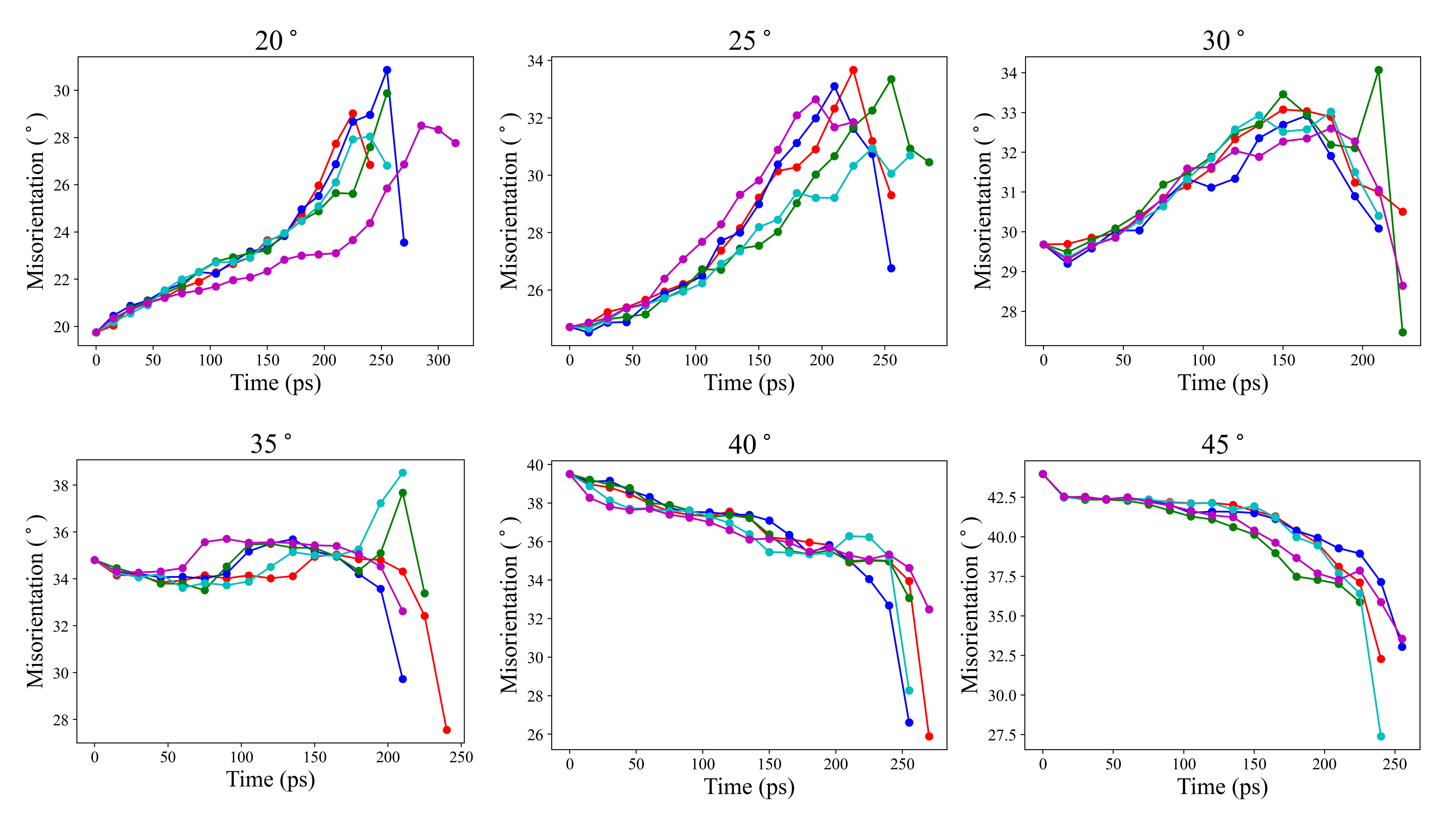}
    \caption{Misorientation vs. time plots for initial misorientations 20-45\degree}
    \label{fig:rotation_curves}
\end{figure}

For each initial misorientation at 1000 K, five different initial velocity seeds are selected for grain rotation measurement using quaternions. The misorientation vs. time plots are summarized in Figure \ref{fig:rotation_curves}. For each initial misorientation, the change in misorientation is significantly affected by initial velocity distribution. For the same initial structure and different initial velocity seeds, the direction of rotation is initially the same but the rotation rates are different. When the grain size becomes relatively small (radius $<$ 18Å), the rotation of the inner grain can be very erratic, with the possibility of moving to both higher and lower misorientations and at various rotation rates. Overall, cylindrical grains with initial misorientations $<35\degree$ rotate towards higher misorientations, and grains with initial misorientations $>35\degree$ rotate towards lower misorientations. For the $\theta = 35\degree$ boundary, very little rotation is observed until the grain becomes small (radius $<$ 18Å). There seems to be a special misorientation angle that all boundaries rotate towards, but never exactly end up at. The same general directions of rotation were also observed by Trautt and Mishin in FCC Cu, and the 36.9$\degree$ misorientation, which corresponds to the $\Sigma5$ boundary, was hypothesized to be the special angle \cite{trautt_grain_2012}.

\begin{figure}[h]
    \centering\leavevmode
    \includegraphics[width=\textwidth]{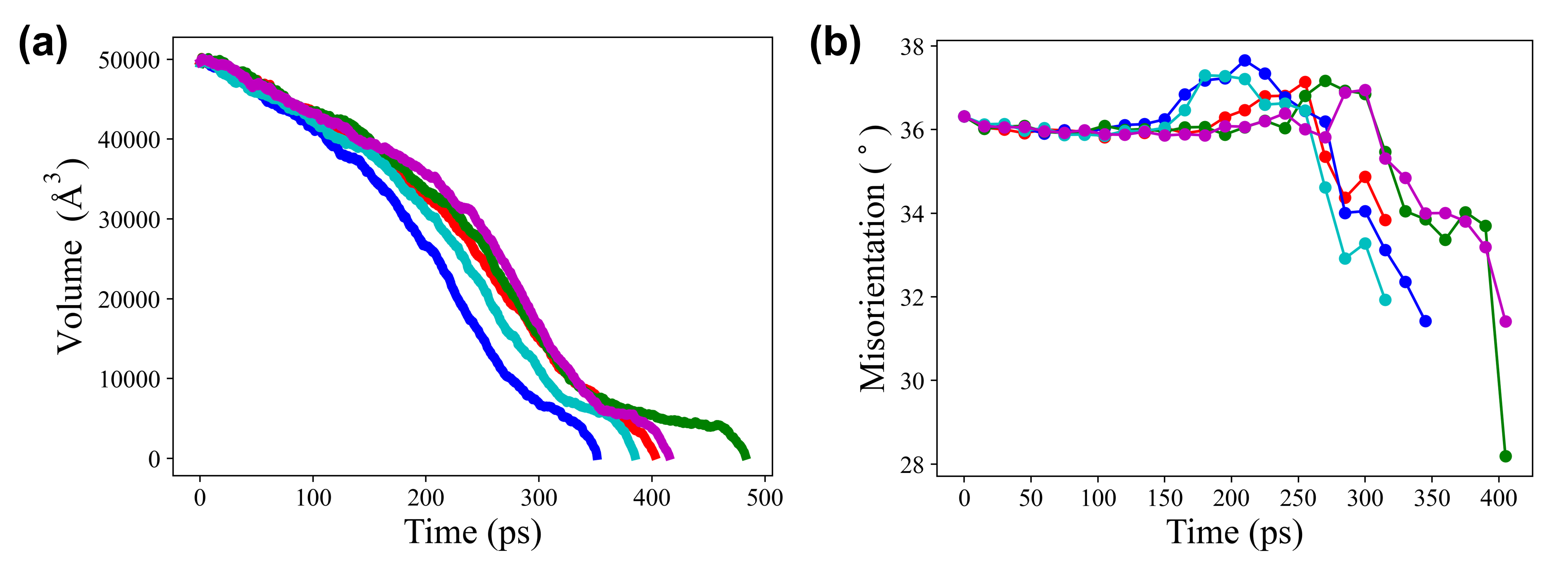}
    \caption{(a) Volume vs. time plot (b) Misorientation vs. time plot for $\Sigma5$ (36.9\degree) boundary}
    \label{fig:sigma5}
\end{figure}

The volume vs. time plots and misorientation vs. time plots of the $\theta = 36.9\degree$ boundary at 1000 K are shown in Figure \ref{fig:sigma5}. The $\theta = 36.9\degree$ boundary is observed to undergo a significant amount of faceting during shrinking, compared with other high angle boundaries at the same temperature. Its migration process is divided into three distinct stages with different migration rates. In the first stage, there is almost no rotation; in the second stage, forward rotation by a small amount is observed; in the third stage, overall backward rotation is observed. This is in agreement with Trautt and Mishin's hypothesis that it does appear to represent a rotation-free boundary misorientation.

\subsection{Temperature}
\begin{figure}[h]
    \centering\leavevmode
    \includegraphics[width=\textwidth]{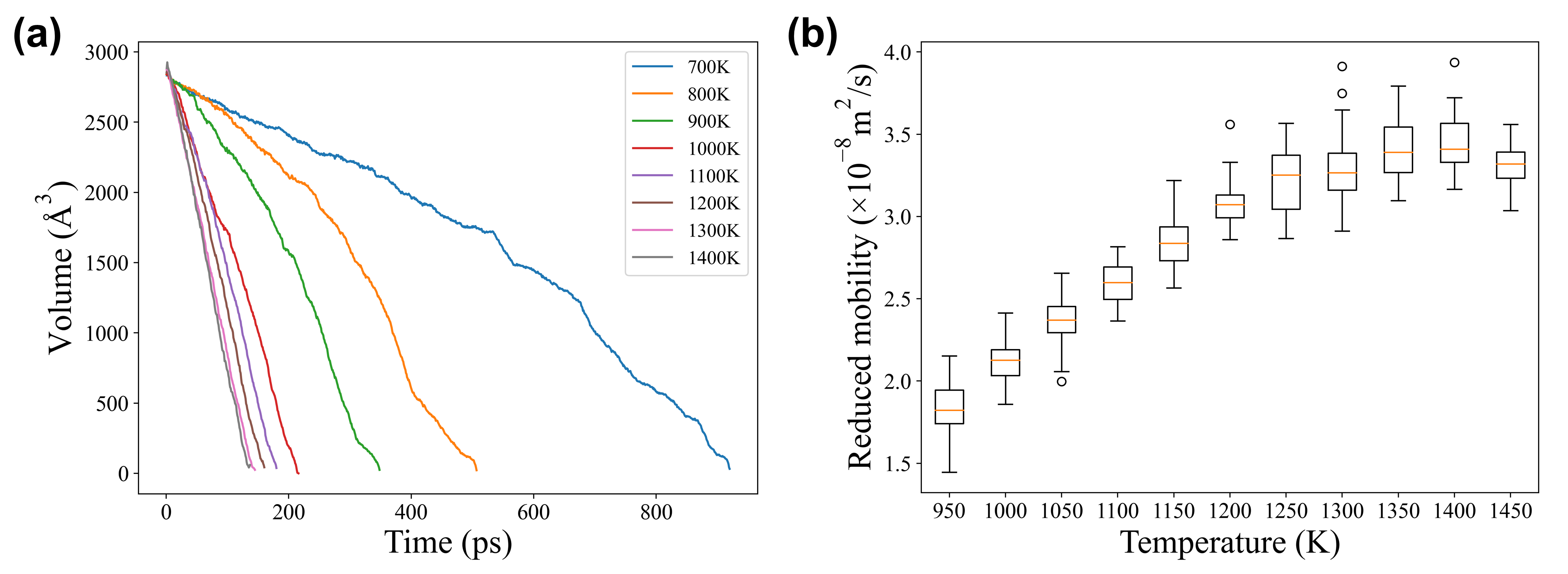}
    \caption{(a) Volume vs. time plots of the $\theta = 30\degree$ free boundary at various temperatures (b) Boxplot of reduced mobility the for $\theta = 30\degree$ free boundary at 950-1450 K}
    \label{fig:mobility_temp_30}
\end{figure}

Temperature is a key factor governing grain boundary migration. In our system, reduced mobility can only be reliably measured for temperatures greater than 900 K. For temperatures $<$ 900 K, the migration rates of the boundaries are constantly changing. Figure \ref{fig:mobility_temp_30}(a) shows the volume vs. time plots for the $\theta = 30\degree$ free boundary at temperatures ranging from 700 to 1400 K. The reduced mobility only becomes constant and measurable in a single run at temperatures above 900 K. A boxplot of reduced mobilities from 950 K to 1450 K is shown in Figure \ref{fig:mobility_temp_30}(b). The average reduced mobility increases linearly from 950 K to 1200 K, non-linearly from 1200 K to 1400 K, and then decreases. For other boundaries, similar trends in reduced mobility can be observed. See Supplementary Information for more details.

\begin{figure}[h]
    \centering\leavevmode
    \includegraphics[width=\textwidth]{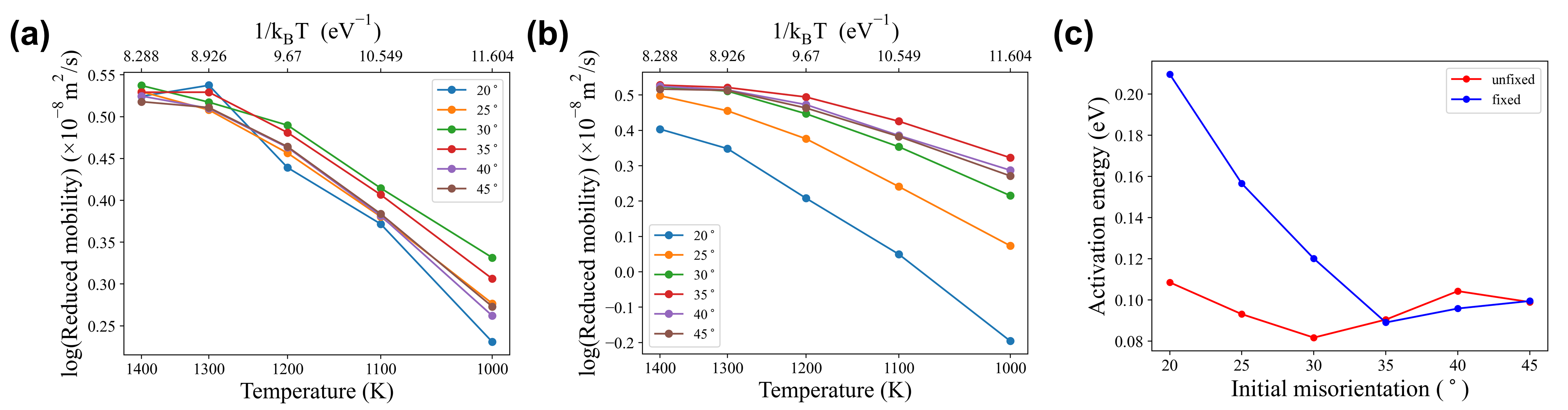}
    \caption{Arrhenius diagrams for (a) free (b) fixed boundaries}
    \label{fig:Arrhenius}
\end{figure}

The average reduced mobilities for the different initial misorientations at different temperatures are summarized in the Arrhenius diagrams in Figure \ref{fig:Arrhenius}(a) and (b). The Arrhenius plots for the freely moving boundaries with different initial misorientations are similar, while for the fixed boundaries, they are more distinct from each other. 

Grain boundary motion is commonly related to thermally activated processes. If grain boundary mobility is  temperature dependent, then it 
should fit to the Arrhenius relationship \cite{turnbull_theory_1951}:

\begin{equation}\label{12}
M = M_0 e^{-\frac{Q}{k_{B} T}} 
\end{equation}

, where $M$ is the grain boundary mobility, $M_0$ is a constant prefactor, $Q$ is the activation energy, $k_B$ is the Boltzmann constant, and $T$ is the temperature. We can assume that the Arrhenius plots are linear for 1000-1200 K, for which the Arrhenius relationship can be applied. The activation energies for the 20-45$\degree$ boundaries are calculated from the Arrhenius plots and shown in Figure \ref{fig:Arrhenius}(c). For the free boundaries, the activation energies for the different initial misorientations are similar, while for the fixed boundaries, they are different. Presumably, for the unfixed boundaries, the activation energies are similar because grain rotation moves the grains toward a common state, and in the fixed boundaries, rotation is prohibited so the grains remain distinct. For the fixed boundaries, the lowest activation energy is found at the $\theta = 35\degree$ boundary, which is the nearest simulated boundary to the $\theta = 36.9\degree$, or $\Sigma5$ boundary. The computed activation energies from molecular dynamics simulations are generally much lower than experimental values. The values computed from our simulations are similar in magnitude to comparable simulations in curved boundaries \cite{zhang_computer_2004,trautt_grain_2012}.

\subsection{System size}
\begin{figure}[h]
    \centering\leavevmode
    \includegraphics[width=\textwidth]{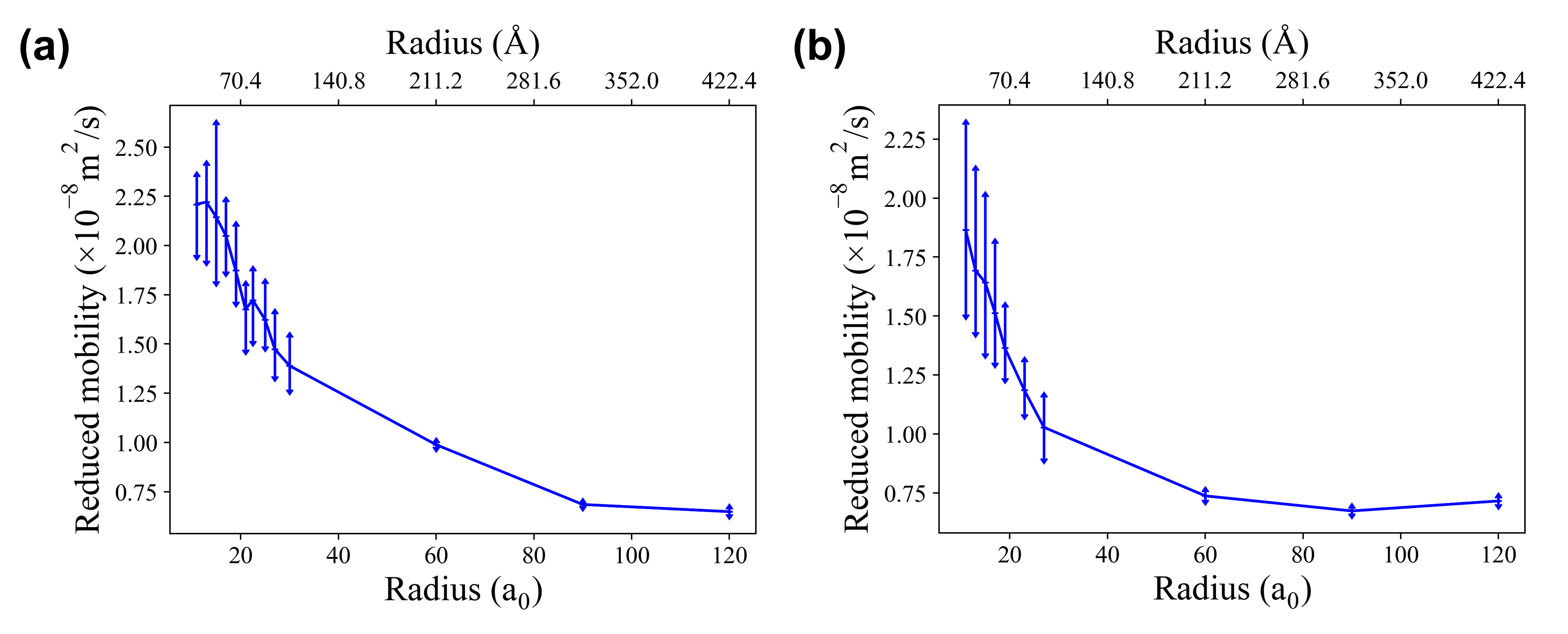}
    \caption{Reduced mobility vs. system size plots for $\theta = 30\degree$ (a) free (b) fixed boundaries at 1000 K}
    \label{fig:sizes}
\end{figure}

The reduced mobilities of cylindrical grain boundaries have been shown to decrease with the increase of system size in half-loop systems[7]. In our simulations, we construct systems of different sizes by varying the radius of the initial cylindrical grain without changing the length of the margin, which is the smallest distance from the outmost point on the cylindrical grain to the side of the simulation box. A set of 30 simulations is performed for each cylindrical bicrystal system with initial radius from 11$a_0$ to 30$a_0$. For systems with initial radius of 60$a_0$, 90$a_0$, and 120$a_0$, we are unable to perform so many simulations due to the limit of available computational resources, so the results are taken from 3 simulations. As shown in Figure \ref{fig:sizes}, with the increase in radius, the average reduced mobility and the range of reduced mobility decrease. At a radius of around 90$a_0$, the reduced mobility starts leveling off with increasing radius. The effects of margin sizes on reduced mobility are also investigated. The reduced mobility is not greatly affected by margin sizes, as long as the margin is not so small that the periodic images start affecting each other, which implies that it is the inner grain size that affects the grain boundary motion rate most significantly.

\section{Mechanistic origins of uncertainty in reduced mobility: dynamic propensity}
\label{sec:analysis}

We propose possible explanations in attempt to explain the mechanisms behind the effect of initial velocity configurations on grain boundary motion.

It has been suggested that the stochastic nature of grain boundary dislocation annihilation could cause variations in grain boundary motion that could lead to differences in measured mobility. Trautt and Mishin examined the dislocation annihilation mechanism and found that it can occur at MD time scales due to dissociation and recombination \cite{trautt_grain_2012}. Furthermore, if grain boundary dislocation annihilation is viewed as the interaction of two random walking dislocations, the distribution of annhilation times should be lognormal, not normal as seen here. Thus, we must look for alternative explanations for the observed variation in measured mobility.

The concept of dynamic propensity was coined by Widmer-Cooper \textit{et al}. to measure the particles' propensity for motion in supercooled liquids \cite{widmer-cooper_how_2004}. The dynamic propensity $p_i$ for the $i$-th particle was originally defined as its mean squared displacement (MSD), or squared displacement averaged over the isoconfigurational ensemble, in a given amount of time:
\begin{equation}\label{propensity}
p_{i} = <\Delta \textbf{r}_i^2>_{IC}
\end{equation}
, where $\Delta \textbf{r}_i$ represents the particle's displacement in the time interval, and $<..>_{IC}$ represents the average over the isoconfigurational ensemble.

\begin{figure}[h]
    \centering\leavevmode    \includegraphics[width=\textwidth]{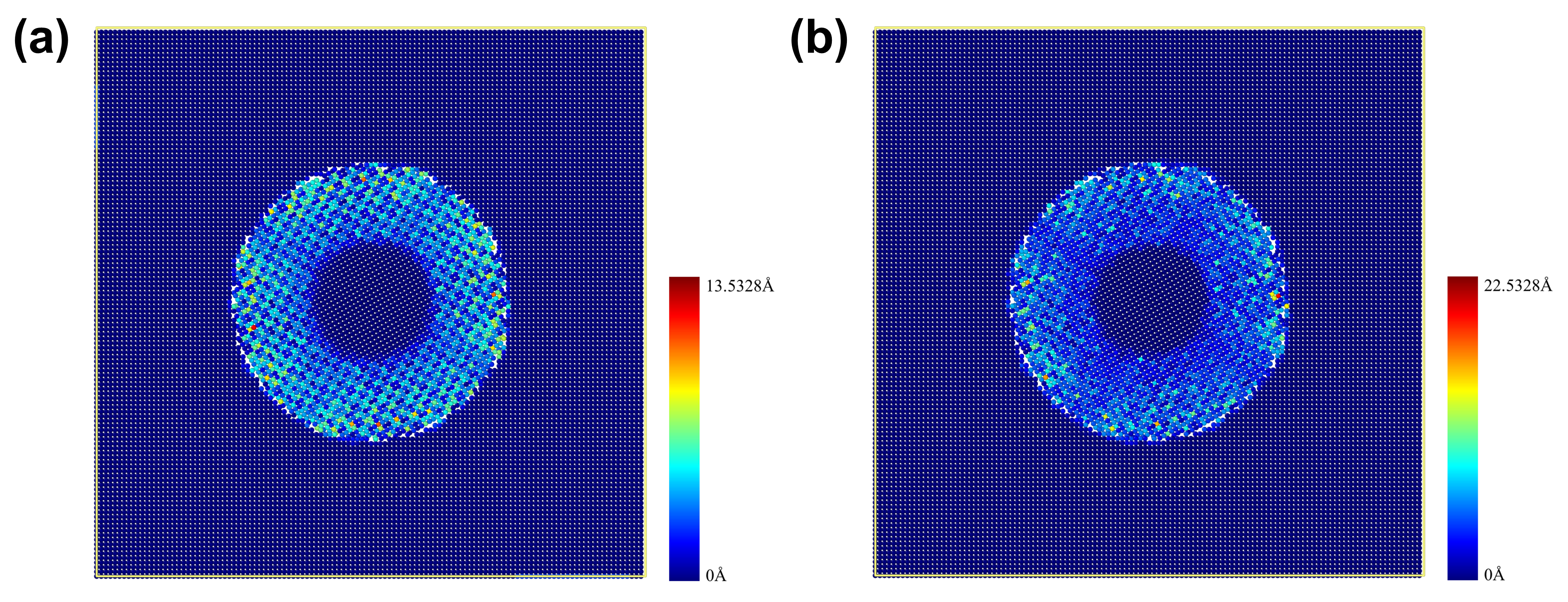}
    \caption{Atomic colormap of the $\theta = 30\degree$ fixed boundary at 1000 K by (a) propensity (b) STD}
    \label{fig:fix-colormap}
\end{figure}

Here we adapt the concept of dynamic propensity to better fit our cylindrical bicrystal system. The displacement $\Delta \textbf{r}_i$ now represents the particle $i$'s displacement between the initial and final system states. The initial system state is obtained by rescaling the atomic coordinates of the energy minimized structure at 0 K by the temperature-dependent lattice constant at the desired temperature, in order to offset the effect of temperature-induced expansion in the NPT ensemble. To keep the analysis simple, we disregard grain rotations and focus on fixed boundaries. In the case of the $\theta = 30\degree$ fixed boundary, the final system state is when the radius of the inner cylindrical grain becomes less than 25Å, so that the distance travelled by the grain boundary is approximately the same in every parallel simulation. In 100 parallel runs with different initial velocity seeds, the displacements of all the atoms between the initial and final states are obtained. The standard deviations of the squared displacements in the parallel simulation runs for all the particles (STD) are calculated to reflect the variability of propensity. The particles' propensities are strongly correlated with their initial positions. The high propensity atoms are arranged in patches in the area travelled through by the grain boundary, with the largest patches clustered around the initial grain boundary, and the high standard deviation atoms are also mostly clustered around the initial grain boundary. For more information regarding the free boundary, refer to Supplementary Information.

\begin{figure}[h]
    \centering\leavevmode
    \includegraphics[width=0.85\textwidth]{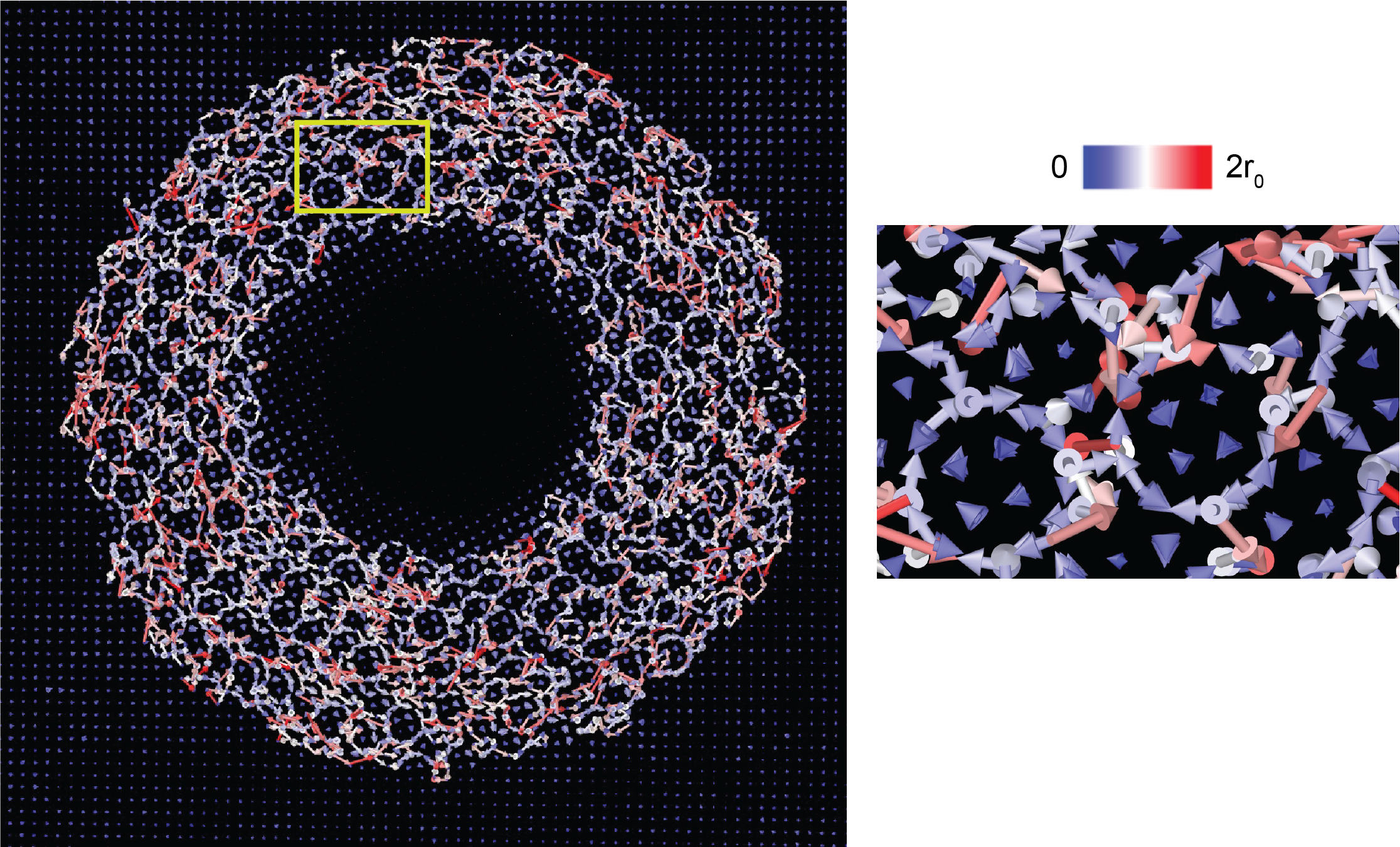}
    \caption{An example net displacement field for the $\theta = 30\degree$ fixed boundary is shown with atomic displacement colored by displacement magnitude ($r_0$ is the first nearest neighbor distance). A subset of the data highlighted in yellow on the left is magnified on the right. Low propensity atoms correspond to atoms which move with short, ordered shuffles.}
    \label{fig:shuffle}
\end{figure}

Figure \ref{fig:shuffle} illustrates a typical displacement field for the constrained migration of the $\theta = 30\degree$ fixed boundary. It is observed that the patches of low propensity atoms in Figure \ref{fig:fix-colormap}(a) correspond to sets of four atoms undergoing short, ordered shuffles during migration. Longer displacements surround these ordered regions of shorter displacements in a cell-like pattern similar to prior work on characterization of constrained migration mechanisms in flat boundaries \cite{chesser_taxonomy_2022}. In analogy to \cite{chesser_taxonomy_2022}, the short shuffles are likely locally distance minimizing displacements in the dichromatic pattern separating nearly coincident sites in the two grains. The longer displacements accommodate poor lattice matching in the dichromatic pattern and involve multiple diffusive hops which can occur in many directions. These long displacements contribute to high propensity. 

\begin{figure}[h]
    \centering\leavevmode
    \includegraphics[width=\textwidth]{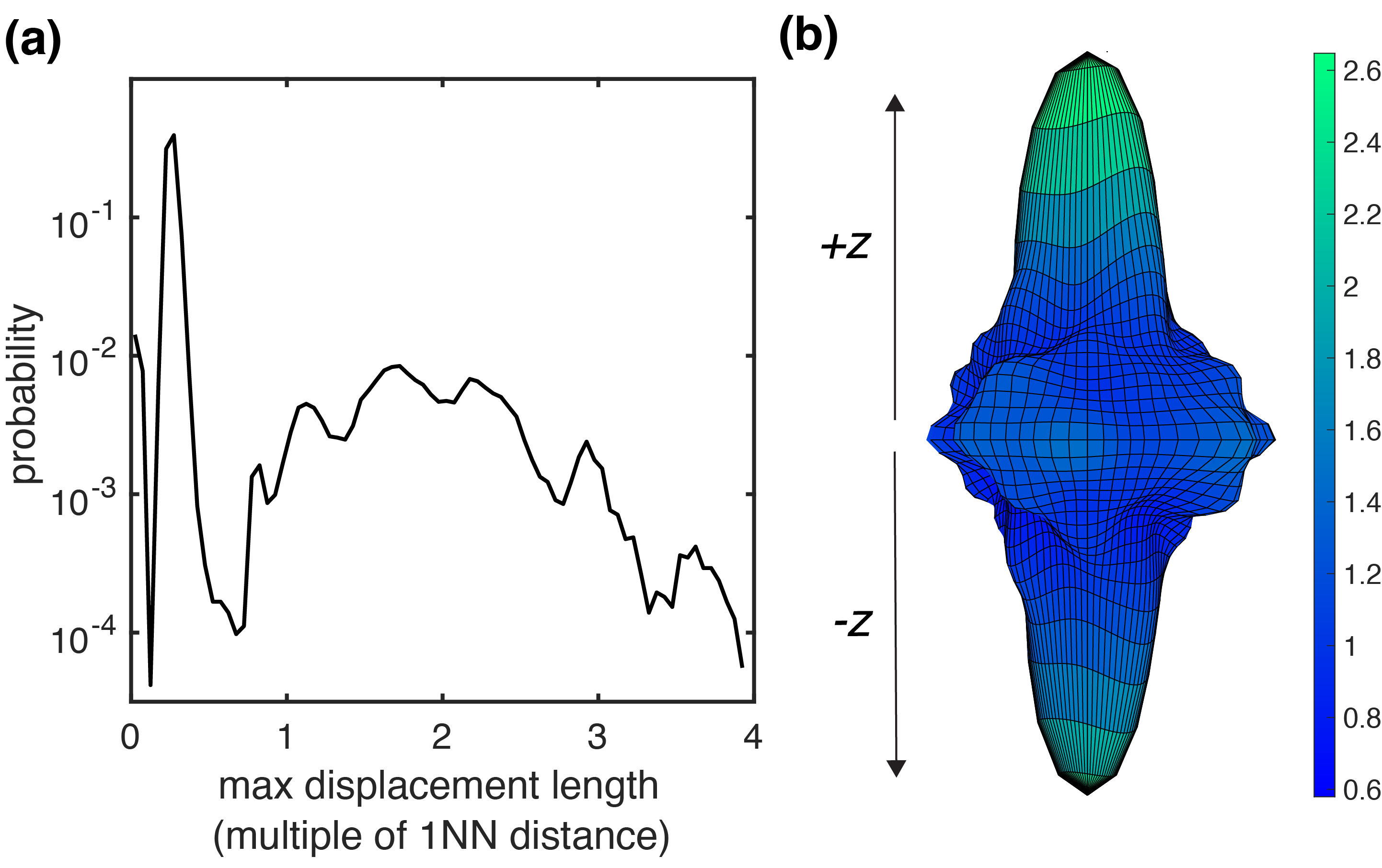}
    \caption{Probability distribution of (a) maximum displacement lengths (b) orientations of maximum displacements weighted by squared net displacement magnitude for the $\theta = 30\degree$ fixed boundary}
    \label{fig:max}
\end{figure}

For individual atoms, different inital velocity seeds lead to motion trajectories of different lengths in various directions. It is found that the highest propensity atoms have propensity values which are dominated by one or a few long diffusion trajectories. The high standard deviations in atoms also result from one or a few relatively long trajectories out of mostly short trajectories. Information regarding the maximum displacements of all the atoms of the $\theta = 30\degree$ fixed boundary from the 100 simulations are summarized in Figure \ref{fig:max}. As shown in Figure \ref{fig:max}, these long displacements have displacement magnitudes up to 4 times the nearest neighbor distance and are most commonly oriented along the tilt axis (z direction), suggesting a pipe diffusion mechanism, which is supported by the existence of multiple threading dislocations in the grain boundary. Although pipe diffusion is not in the direction of GB motion, it may play an important role in stimulating or suppressing normal migration. For example, if we assume that normal migration is mediated by disconnections, then pipe diffusion provides a point defect flux through disconnection cores which can modify local disconnection migration barriers. 

We speculate that the wide range of reduced mobilities stem from the creation of multiple metastable states in the initial stage of the simulation. The random assignment of initial velocities takes grain boundary atoms to different positions, which leads to a variety of metastable grain boundary states with different energies that have different migration rates \cite{han_grain-boundary_2016}. Pipe diffusion, in particular, can take atoms far away from their initial positions in the initial stage, resulting in long displacements in the z direction, and therefore high propensity. In the initial stage of the simulation, when the grain boundary area is largest and the number of threading dislocations is highest, there is higher chance for grain boundary atoms to travel longer distances through pipe diffusion, leading to high propensity atoms occurring most frequently in the beginning of motion. Point defect-disconnection interactions as well as metastable states should be studied in more depth in the future to provide further insights into the wide variability of migration rates with initial velocity seed. 

\section*{Conclusions}
\label{sec:conclusions}

Grain boundary motion is greatly affected by the initial velocity distribution when all other conditions remain the same, in contrary to the popular belief that grain boundary motion is mostly deterministic. In order to investigate the effects of initial velocity distributions on grain boundary migration, we implement many parallel simulations with different initial velocity distributions in the isoconfigurational ensemble for various cylindrical bicrystal systems. For cylindrical grain boundaries capable of undergoing steady-state motion, the reduced mobilities resulting from different initial velocity seeds in LAMMPS form wide-ranging Gaussian distributions with a range of around $\pm 20\%$ of the mean value, indicating a high uncertainty in reduced mobility measurement. In order to identify the source of the uncertainty, we create atomic colormaps to highlight atoms with high propensity for motion, and atoms that are most affected by initial velocity seeds, both of which are highly correlated with their initial positions. High propensity and high standard deviations are often the result of one or a few very long diffusive trajectories. The maximum displacements can have magnitudes as large as 4 times the nearest neighbor distance, and have the highest probability of being oriented in the direction of the tilt axis, suggesting a pipe diffusion mechanism, which does not contribute directly to mobility, but can have a great impact on mobility by changing the pathway and energy barrier of grain boundary migration.

The large variation in the measured mobilities for a given boundary crystallography changes the way we view grain boundary motion. Grain boundary mobility is no longer seen as an intrinsic property of a grain boundary, but a spectrum of values that result from the different assignments of initial velocities.

\section*{Acknowledgments}

Work at CMU was supported by National Science Foundation grant DMR-1710186.

\newpage



\end{document}


\begin{frontmatter}

\title{Supplementary Information}
\author[cmu]{Anqi Qiu}
\ead{anqiq@andrew.cmu.edu}

\author[gmu]{Ian Chesser}
\ead{ichesser@andrew.cmu.edu}

\author[cmu]{Elizabeth Holm}
\ead{eaholm@cmu.edu}

\address[cmu]{Department of Materials Science and Engineering, Carnegie Mellon University, Pittsburgh, PA 15213, USA}
\address[gmu]{Department of Physics \& Astronomy, George Mason University, Fairfax, VA 22030}
\end{frontmatter}

\section{Grain rotation visualization}
\begin{figure}[ht!]
    \centering\leavevmode
    \includegraphics[width=0.9\textwidth]{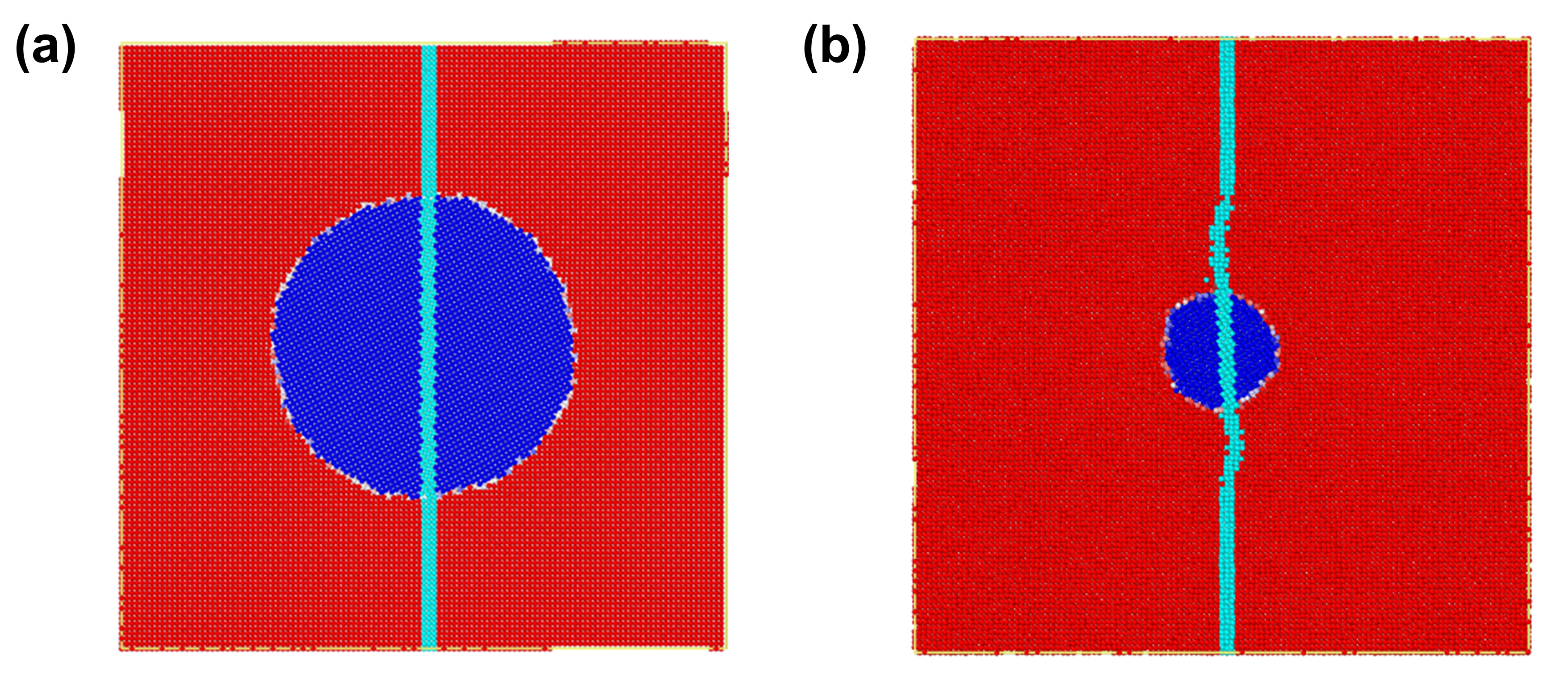}
    \caption{(a) Initial state (b) rotated grain shown by deformed fiducial marker}
    \label{fig:fiducial}
\end{figure}

 The rotation of the cylindrical grain can be visualized with a fiducial marker that initially runs straight through the center of the simulation box, as shown in light blue in Figure \ref{fig:fiducial}. As the cylindrical inner grain rotates, the marker is deformed, indicating the direction and extent of rotation. As the grain boundary migrates, the atoms in the inner grain are swept out by the motion of the grain boundary and become part of the outer grain.  

The rotations of the inner cylindrical grains with different initial misorientations at 1000K are visualized in Figure \ref{fig:rotations_fiducial}, with the atoms swept out by the grain boundary made transparent. Grain rotation occurs for all of the initial misorientations shown. The fiducial marker is unable to measure rotation accurately and efficiently, so quaternions are used for the accurate measurement of grain rotation. The measurement of grain rotation using quaternions matches the measurement using fiducial markers.

\begin{figure}[ht!]
    \centering\leavevmode
    \includegraphics[width=0.8\textwidth]{Elsevier_paper_template/Figures/rotations_fiducial.jpg}
    \caption{Grain rotations in high angle boundaries at 1000K visualized with fiducial markers}
    \label{fig:rotations_fiducial}
\end{figure}

\section{Temperature}
\begin{figure}[ht!]
    \centering\leavevmode
    \includegraphics[width=\textwidth]{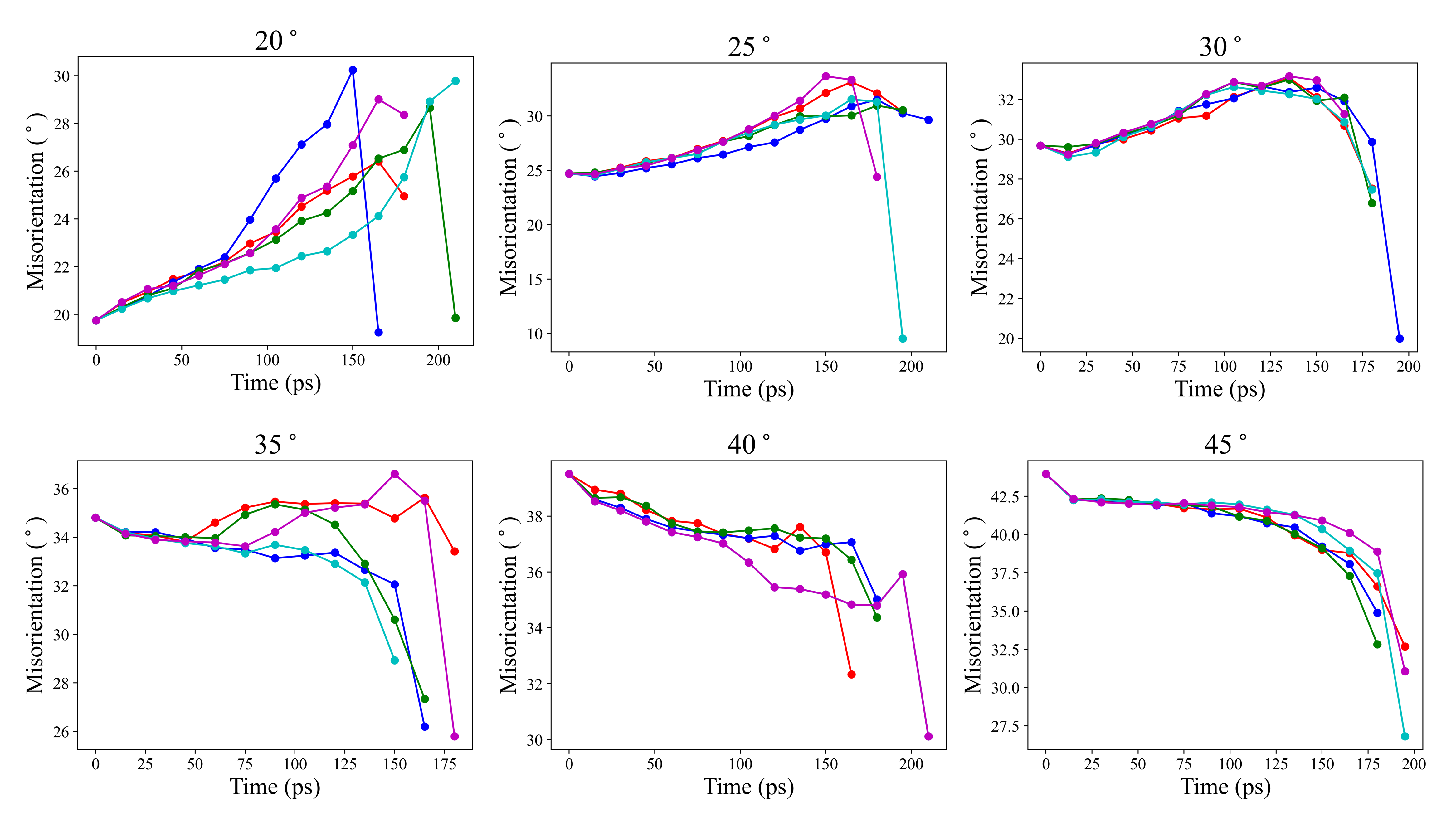}
    \caption{Grain rotations for 20-45$\degree$ boundaries at 1100K}
    \label{fig:rotation-1100K}
\end{figure}

\begin{figure}[ht!]
    \centering\leavevmode
    \includegraphics[width=\textwidth]{   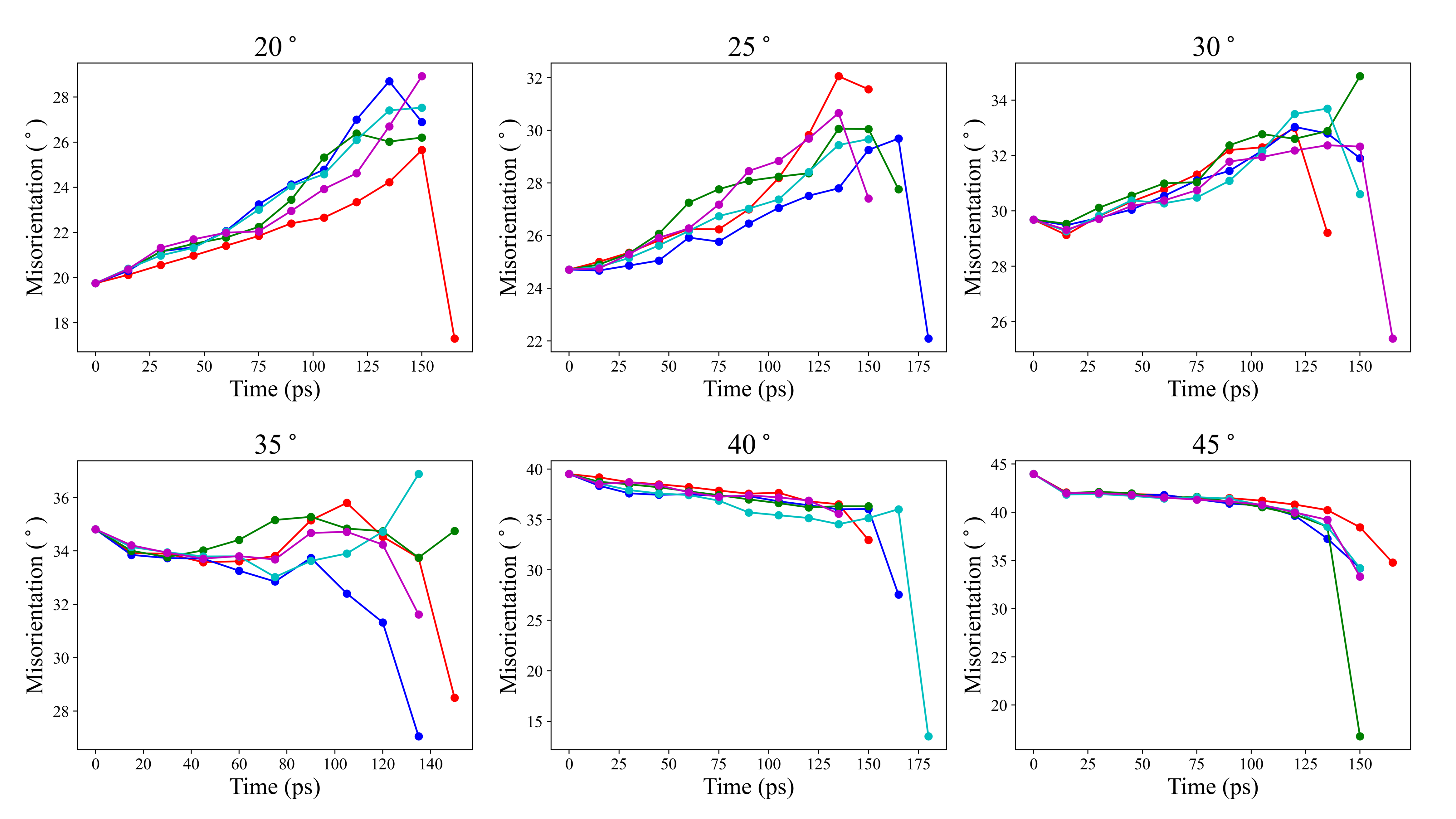}
    \caption{Grain rotations for 20-45$\degree$ boundaries at 1200K}
    \label{fig:rotation-1200K}
\end{figure}

\begin{figure}[ht!]
    \centering\leavevmode
    \includegraphics[width=\textwidth]{   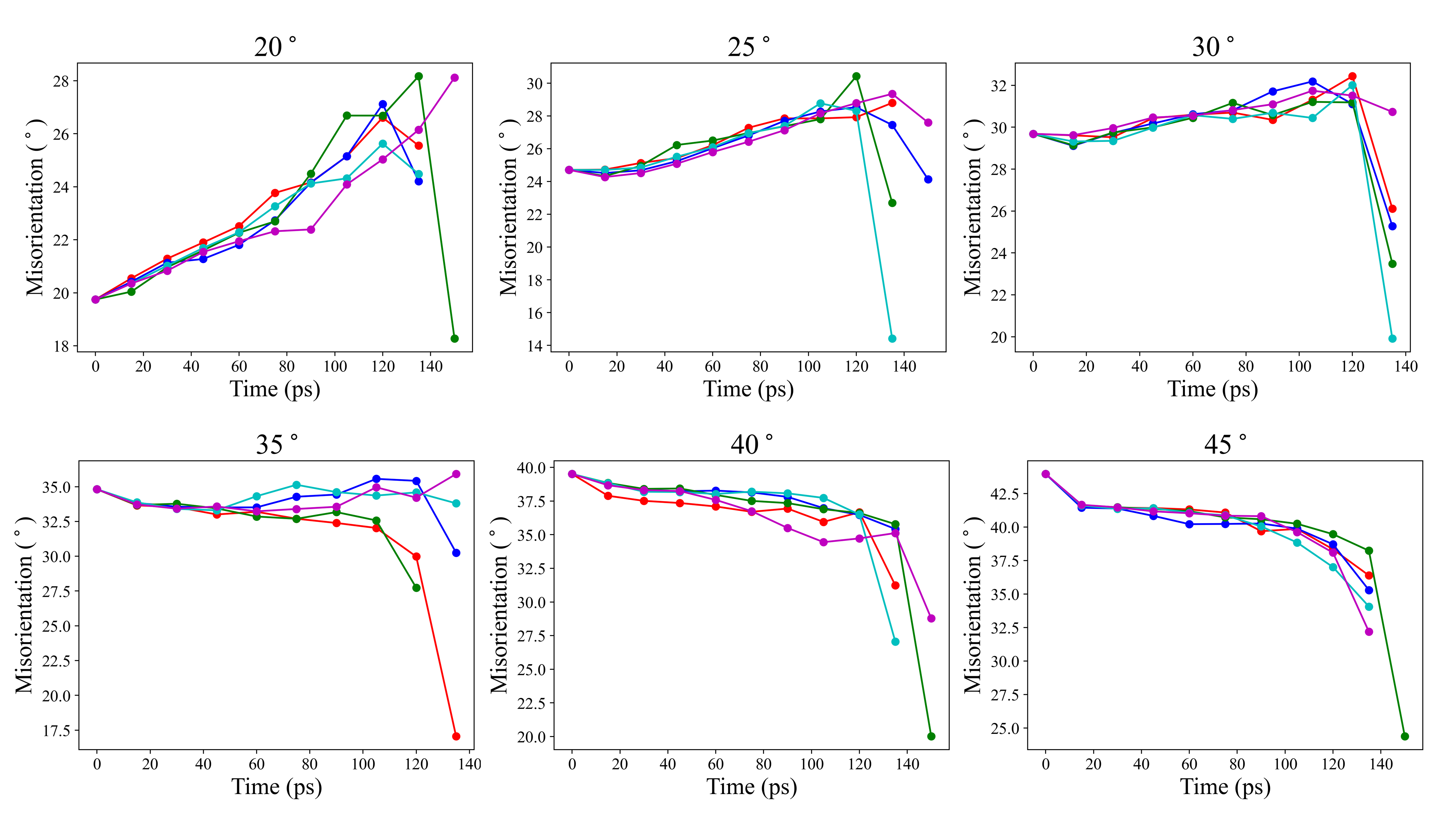}
    \caption{Grain rotations for 20-45$\degree$ boundaries at 1300K}
    \label{fig:rotation-1300K}
\end{figure}

\begin{figure}[ht!]
    \centering\leavevmode
    \includegraphics[width=\textwidth]{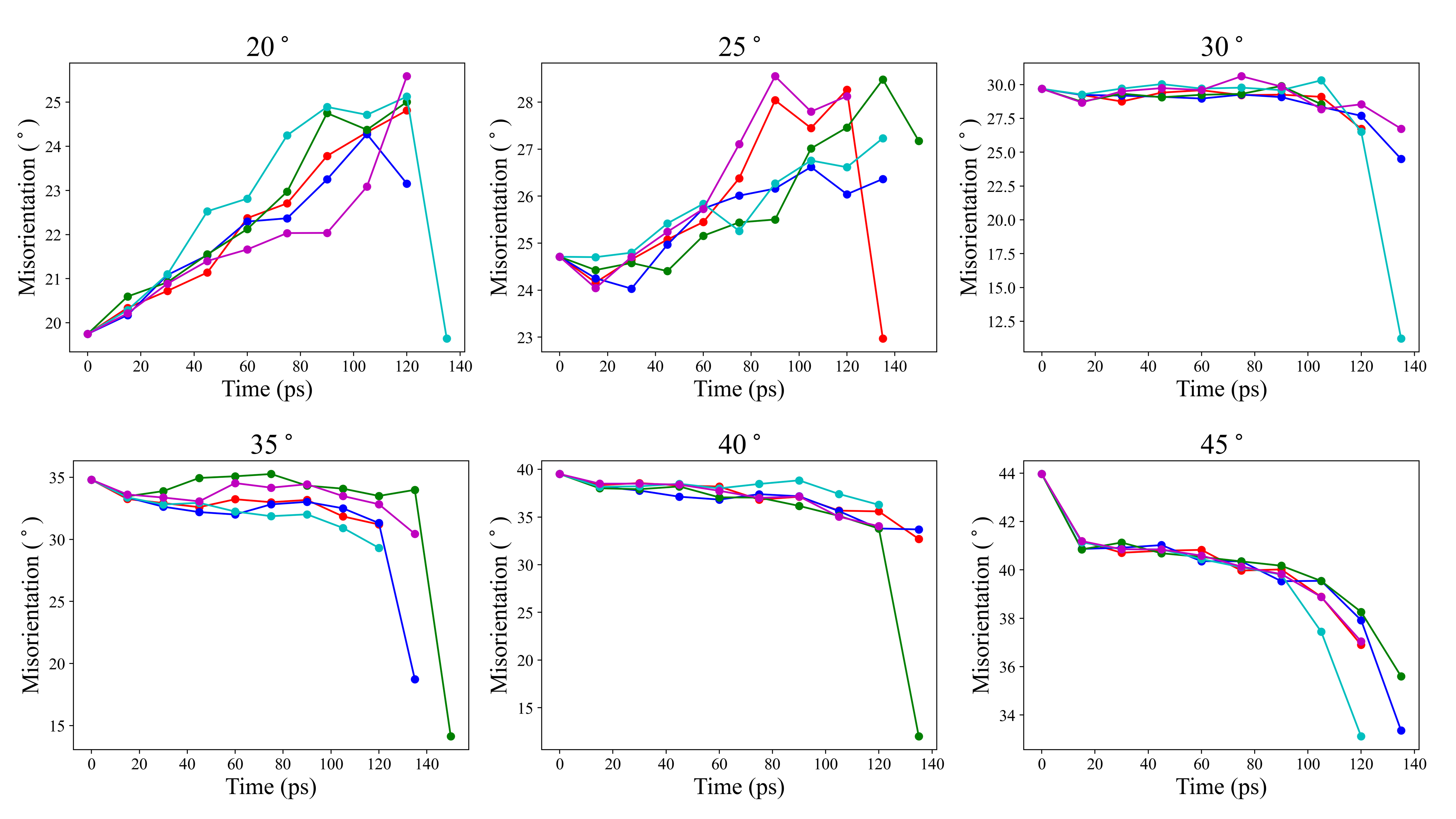}
    \caption{Grain rotations for 20-45$\degree$ boundaries at 1400K}
    \label{fig:rotation-1400K}
\end{figure}

The reduced mobility vs. temperature plots for initial misorientations of 20-45$\degree$ at 1100-1400K are shown in Figure \ref{fig:rotation-1100K} - \ref{fig:rotation-1400K}. The amount of grain rotation is reduced at higher temperatures, while the directions remain the same, possibly due to grain boundary pre-melting weakening the shear-coupling between the inner and outer grains. Grain rotations are significantly affected by initial velocity seeds at higher temperatures.

 For the 20-45$\degree$ boundaries, the reduced mobility increases linearly for 1000-1200K, and then non-linearly for 1200-1400K. For the 25$\degree$ and 25$\degree$ boundaries, the average reduced mobilities of the fixed boundaries are always lower than their free counterparts at the same temperature. For the 30$\degree$ boundaries, the average reduced mobilities of the free and fixed boundaries become closer with increasing temperature. For the 35-45$\degree$ boundaries, the average reduced mobilities for the fixed and free boundaries are very similar, especially for 45$\degree$, though the ranges are different. 
 
\clearpage
 \begin{figure}[ht!]
    \centering\leavevmode
    \includegraphics[width=\textwidth]{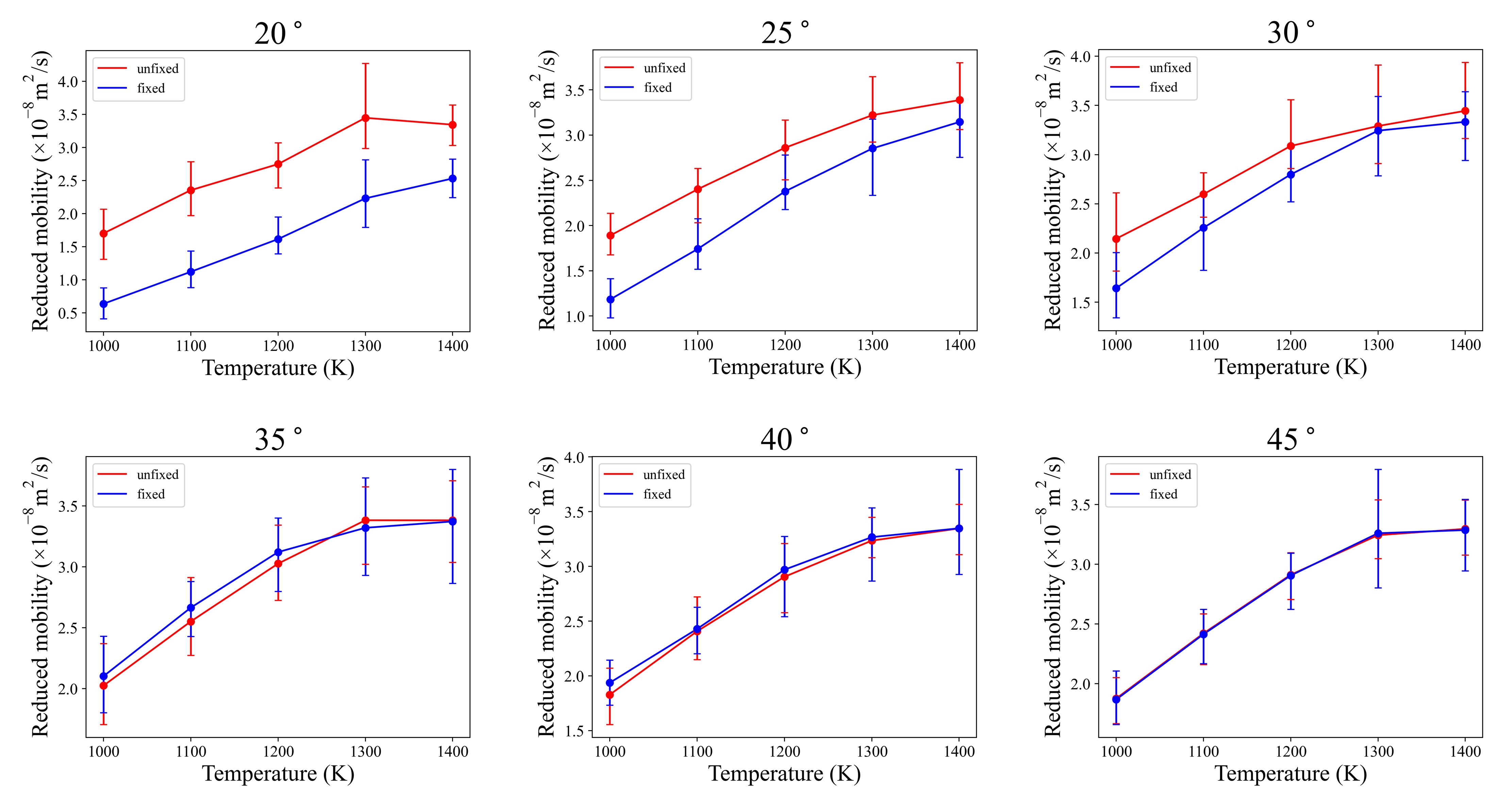}
    \caption{Reduced mobility vs. temperature plots for 20-45$\degree$ boundaries}
    \label{fig:mobility_temps}
\end{figure}

\clearpage
\newpage
\section{Low angle grain boundaries}
Low angle grain boundaries ($\theta<=15 \degree$) consist of arrays of well-defined and widely separated dislocations, unlike high angle grain boundaries in which atoms are more connected. Due to high energy anisotropy, low angle curved [001] tilt boundaries are unable to consistently assume cylindrical shapes during shrinkage and undergo steady-state motion\cite{verhasselt_shape_1999,kirch_inclination_2008}. For boundaries with initial misorientations $< 15\degree$, cylindrical shapes cannot be formed at any point in the simulation. Starting at around 15$\degree$ initial misorientation, the boundary is able to maintain a relatively round shape during motion, though the shape is not consistently cylindrical as multiple cusps and facets are constantly forming and disappearing. The initial shapes of the $\theta = 13\degree$ and $\theta = 15\degree$ boundaries are shown in Figure \ref{fig:13-15}. 
\begin{figure}[h]
    \centering\leavevmode
    \includegraphics[width=0.9\textwidth]{Elsevier_paper_template/Figures/13-15.jpg}
    \caption{Initial shapes of the $\theta = 13\degree$ and  $\theta = 15\degree$ boundaries at 0K}
    \label{fig:13-15}
\end{figure}

\begin{figure}[ht]
    \centering\leavevmode
    \includegraphics[width=0.6\textwidth]{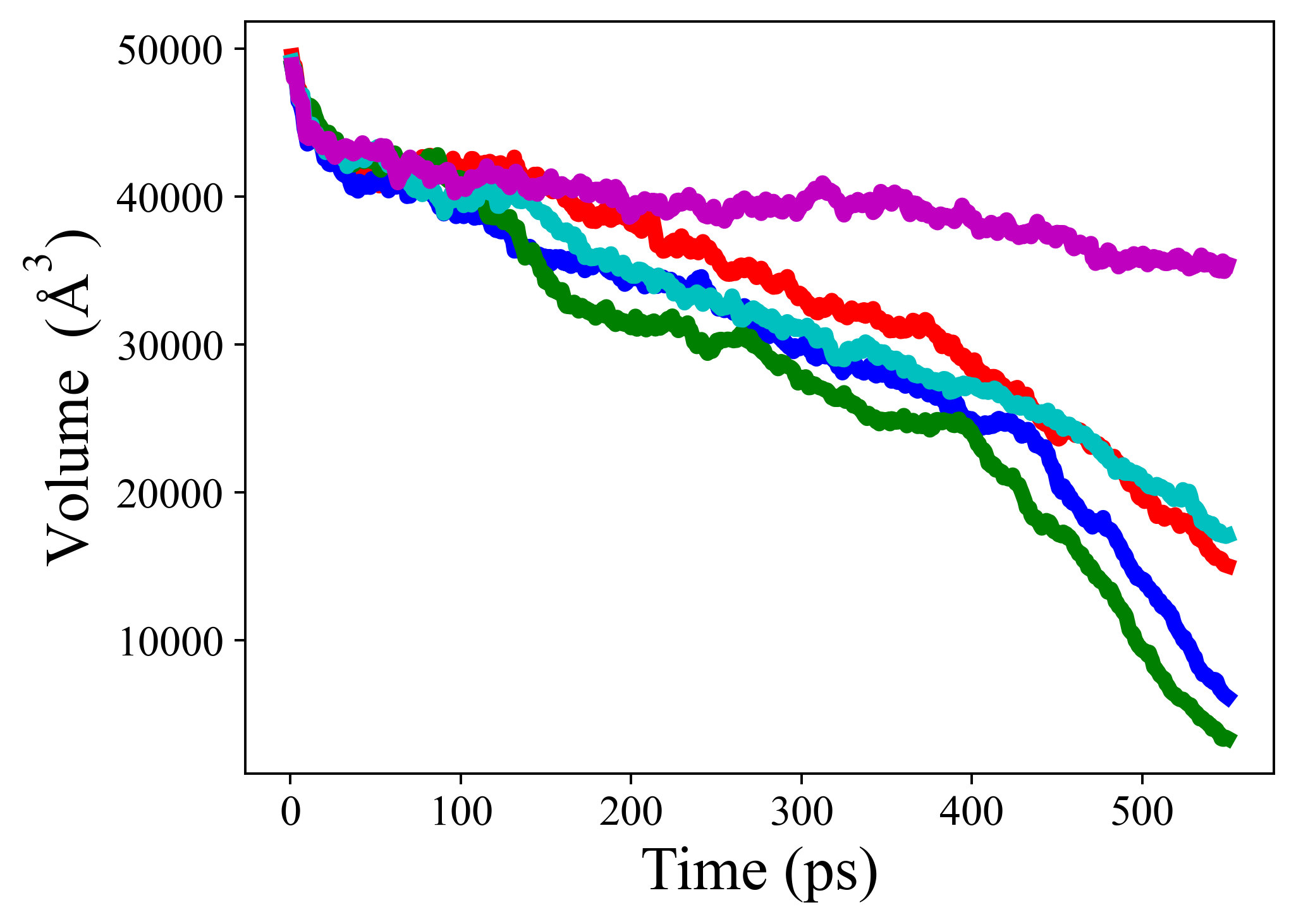}
    \caption{Volume vs. time plots for $\theta = 15\degree$ free boundary at 1000K.}
    \label{fig:low}
\end{figure}

Similar to the $\Sigma5$ boundary, the $\theta = 15\degree$ boundary can have varying reduced mobilities over the course of a single simulation run, and the shapes of the volume vs. time plots vary significantly between runs. Different stages with constant reduced mobilities, stagnant stages with no shrinking, and jagged motions are observed in the different simulation runs. The volume vs. time plots from 5 different simulation runs for the $\theta = 15\degree$ free boundary at 1000K are shown in Figure \ref{fig:low}. These plots are cherry-picked to represent the diversity of the motion in low angle cylindrical grain boundaries. Low angle grain boundaries have the tendency to undergo stages of stagnation, due to vacancy generation, dislocations, faceting, etc\cite{upmanyu_atomistic_1998}, which complicate their motions. The 20$\degree$ and 25$\degree$ boundaries also display some degree of jagged motion and very short periods of stagnation during shrinking, though their overall volume vs. time curves are sufficiently linear and a single mobility value can be extracted from a single run.

\clearpage

\clearpage
\newpage
\section{Dynamic propensity}
\begin{figure}[h]
    \centering\leavevmode    \includegraphics[width=\textwidth]{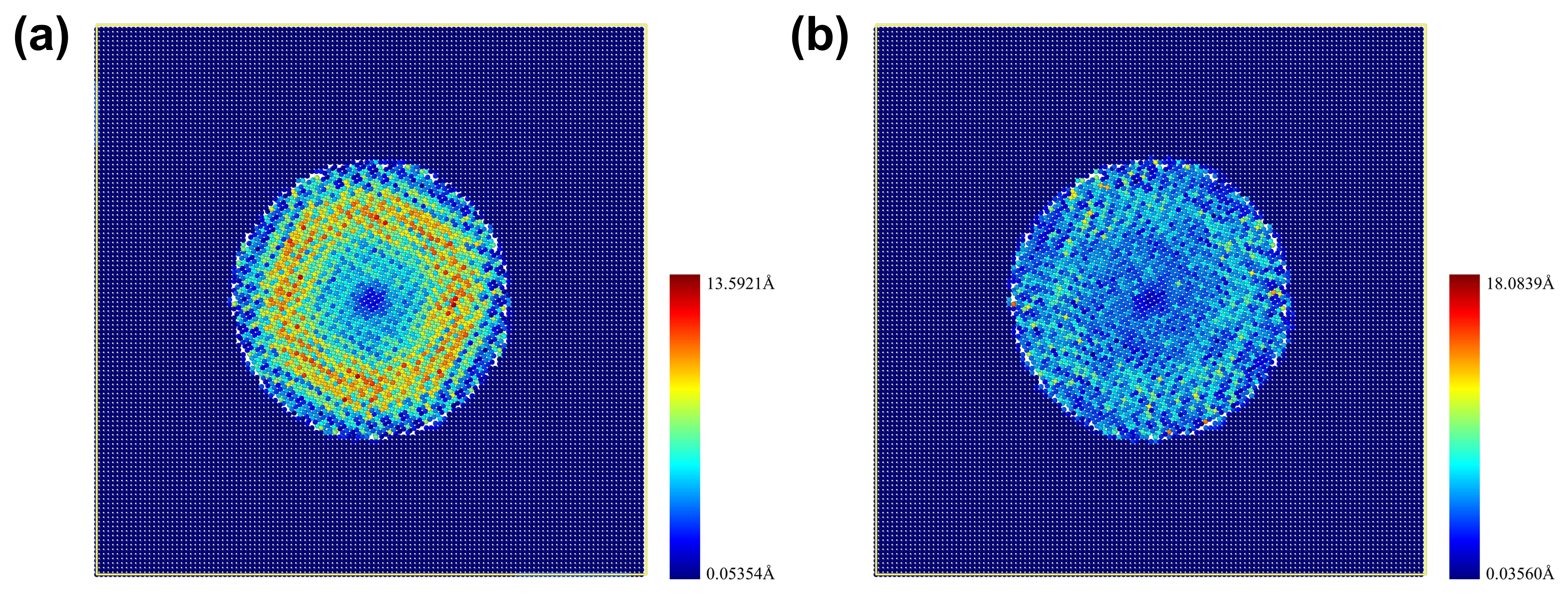}
    \caption{Atomic colormap of the $\theta = 30\degree$ free boundary at 1000K by (a) Propensity (b) STD}
    \label{fig:unfix-colormap}
\end{figure}

The propensity analysis of the freely moving boundary is complicated by displacement vectors related to grain rotation. For the freely shrinking $\theta = 30\degree$ boundary, the colormaps of atomic dynamic propensity and standard deviation are shown in Figure \ref{fig:unfix-colormap}. The atoms with the highest propensities are located some distance away from the initial grain boundary, and the atoms with the highest standard deviations are scattered throughout the inner grain, unlike its fixed counterpart where the high propensity and high standard deviation atoms are clustered around the initial grain boundary. 

The differences in the positions of the high propensity and high standard deviation atoms in the fixed and free boundaries can be attributed to grain rotation, which accumulates in the initial stage.

\bibliographystyle{elsarticle-num}

\newpage
\bibliography{references}